\documentclass[a4paper,12pt]{article}
\usepackage[utf8]{inputenc}
\usepackage{placeins}
\usepackage{amssymb,amsmath,amsfonts}
\usepackage[colorlinks,linkcolor=purple,citecolor=teal]{hyperref}
\usepackage{dsfont}
\usepackage{color}
\usepackage{graphicx}
\usepackage{hyphenat}
\usepackage{wrapfig}
\usepackage{empheq}
\usepackage{textcomp}
\usepackage[caption=false]{subfig}
\usepackage{rotating}
\usepackage{wrapfig}
\usepackage{pdfpages}
\usepackage{verbatim}
\usepackage{setspace}
\usepackage{array}
\usepackage{caption}
\usepackage[pdf]{pstricks}
\usepackage{upgreek}
\usepackage{cmll}
\usepackage{latexsym}
\usepackage{braket}
\usepackage{cite}
\usepackage{epsfig}
\usepackage[left=2.4cm,top=3.3cm,right=2.4cm,bottom=3.3cm,bindingoffset=0cm]{geometry}
\usepackage[titletoc,page]{appendix}

\newcommand{\beq}{\begin{eqnarray}}
\newcommand{\eeq}{\end{eqnarray}}
\newcommand{\bea}{\begin{eqnarray}}
\newcommand{\eea}{\end{eqnarray}}
\newcommand{\be}{\begin{equation}}
\newcommand{\ee}{\end{equation}}

\def\de{\partial}

\def\1{\mathbbm{1}}

\def\a{\alpha}

\def\nn{\nonumber}

\def\vx{\vec{x}}
\def\vt{\vec{\tau}}
\def\xr{\xi^2+\rho^2}
\def\xx{\xi^2}
\def\rr{\rho^2}
\def\vc{\vec{\chi}}
\def\xc{\left(\hat{r}\cdot\vec{\chi}\right)}
\def\xt{\left(\hat{r}\cdot\vec{\tau}\right)}
\def\ct{\left(\vec{\chi}\cdot\vec{\tau}\right)}

\def\abc{\epsilon^{abc}\chi^a\hat{r}^b \tau^c }
\def\eiab{\epsilon^{iab}}

\def\mM{\mathcal{M}}
\def\A{\textbf{a}}
\def\Tr{\qopname\relax o{Tr}}
\def\tr{\qopname\relax o{tr}}
\def\Vol{\qopname\relax o{Vol}}
\def\KK{\qopname\relax o{KK}}
\def\YM{\qopname\relax o{YM}}
\def\CS{\qopname\relax o{CS}}
\def\Nc{{N_{\rm c}}}
\def\Nf{{N_{\rm f}}}
\def\calDn{\mathcal{D}_{\rm n}}
\def\calDp{\mathcal{D}_{\rm p}}
\def\calDD{\mathcal{D}_{\rm D}}
\def\calDN{\mathcal{D}_{\rm N}}
\def\DeltaN{\Delta_{\rm N}}
\def\DeltaD{\Delta_{\rm D}}
\def\dD{d_{\rm D}}

\numberwithin{equation}{section}
\newcommand{\hA}[0]{\widehat{A}}
\newcommand{\hF}[0]{\widehat{F}}
\newcommand{\mF}[0]{\mathcal{F}}
\newcommand{\mA}[0]{\mathcal{A}}

\begin{document}

\title{
\begin{flushright}\ \vskip -1.5cm {\footnotesize {IFUP-TH-2019}}\end{flushright}
\vskip 20pt
\bf{    Deuteron electric dipole moment from holographic QCD
  }
}
\vskip 30pt  
\author{ Lorenzo Bartolini$^{(1)}$,  Stefano Bolognesi$^{(1)}$ and Sven Bjarke Gudnason$^{(2)}$ \\[13pt]
{\em \footnotesize
$^{(1)}$Department of Physics ``E. Fermi", University of Pisa and INFN, Sezione di Pisa
}\\[-5pt]
{\em \footnotesize
Largo Pontecorvo, 3, Ed. C, 56127 Pisa, Italy
}\\[2pt]
{\em \footnotesize
$^{(2)}$Institute of Contemporary Mathematics, School of
  Mathematics and Statistics,}  \\[-5pt]
{\em \footnotesize Henan University, Kaifeng, Henan 475004,
  P.~R.~China}
 \\ [5pt] 
{ \footnotesize lorenzobartolini89(at)gmail.com, stefano.bolognesi(at)unipi.it, gudnason(at)henu.edu.cn}  
}

\vskip 6pt
\date{January 2020}
\maketitle
\vskip 0pt

\begin{abstract}

We compute the electric dipole moment (EDM) of the deuteron in the holographic QCD model of Witten-Sakai-Sugimoto.
Previously, the leading contribution to the EDM of nucleons was
computed, finding opposite values for the proton and the neutron which
then cancel each other in the deuteron state.  
Here we compute the next-to-leading order contribution which provides
a splitting between their absolute value. 
At large $\Nc$ and large 't Hooft coupling $\lambda$, nuclei are
bound states of almost isolated nucleons. 
In particular, we find that in this limit the deuteron EDM is given by
the splitting between proton and neutron EDMs. 
Our estimate for the deuteron EDM extrapolated to the physical values
of $\Nc$, $\lambda$, $M_{\KK}$ and $m_q$ is $d_d = -0.92 \times 10^{-16} \theta\ e \cdot {cm}$.
This is consistent, in sign and magnitude, with results found
previously in the literature and obtained using completely different
methods.

\end{abstract}
\newpage
\tableofcontents

\section{Introduction}

The action of QCD can be supplemented with a topological $\theta$-term
without spoiling its gauge and Lorentz invariance: this term however
introduces CP-violation in the theory, as it can be regarded as an
analog of the $\vec{B}\cdot\vec{E}$ term in electromagnetism. 

The most studied CP-violating observables arising from this term are
electric dipole moments of baryons, $\mathcal{D}_B$, that are linear in
the $\theta$-parameter. Until recent years, following the pioneering
work of Ramsey and Purcell in 1950 \cite{Purcell:1950zz}, most efforts
were directed at predicting the electric dipole moment of the neutron
$\calDn$, which was the most accessible one using direct 
measures: Experimentally an upper bound amounting to
$|\calDn|<3.0\times 10^{-26}e \cdot cm $ has been established
for this observable \cite{Afach:2015sja}, while most estimates set the
value of the $\theta$-induced contribution to the dipole moment to
about $10^{-16}\theta e\cdot cm$. This implies a somewhat unnatural
smallness for the $\theta$ parameter, which is then set to less than
about $\theta \lesssim 10^{-10}$. This unnaturally small, but
eventually nonvanishing amount of CP-violation goes under the name of
the ``strong CP-problem''. 

For the deuteron, the state-of-the-art of the electric dipole moment
$\calDD$ is less rich at the moment. On the experimental
side especially there are no direct measures due to the fact that it is
electrically charged, making it unfit for measurements which involve
placing it in electric fields. Theoretical estimates are essentially
obtained through QCD sum rules \cite{Lebedev:2004va,Liu:2004tq} and
via models of nuclear potential \cite{Yamanaka:2015qfa}\footnote{For a
  review on the topic of EDMs of light nuclei, see
  \cite{Yamanaka:2016umw} 
}
: the tool that provided most estimates for $\calDn$, the chiral
Lagrangian, tends to produce 
electric dipole moments that are equal in magnitude and opposite in
sign for the neutron and the proton, so that the single nucleon
contributions, which are expected to be important, tend to cancel
each other inside the deuteron: nevertheless some results in this
context are available for the $\theta$ induced EDM as lower bounds
\cite{deVries:2011re,deVries:2011an}, while two-nucleon terms
can also be computed \cite{Bsaisou:2012rg,Bsaisou:2014zwa}.

In recent years both the experimental and theoretical
fields have acquired new tools to tackle the problem of the
determination of $\calDD$. On the experimental side, the
development of storage-ring technology allows to measure the electric
dipole moment of charged particles with relevant precision: The
JEDI\footnote{Website: \href{http://collaborations.fz-juelich.de/ikp/jedi/}{http://collaborations.fz-juelich.de/ikp/jedi/}}
collaboration in J\"ulich has a goal of reaching a potential
sensitivity of $10^{-29}\text{e}\cdot\text{cm}$ \cite{Pretz:2013us},
so that there is the possibility, if good theoretical predictions are
available, that the strong CP-problem can be pushed to even more
restrictive regimes, lowering the upper bound on $\theta$. The other
possibility is that instead the experiments find a finite value for
$\calDD$, in which case it would be of paramount importance to
have a quantitatively meaningful theoretical estimate, to infer the
value of $\theta$. 

On the theory side instead, the holographic model of
Witten-Sakai-Sugimoto (WSS)
\cite{Witten:1998zw,Sakai:2004cn,Sakai:2005yt} has been used to
successfully compute the electric dipole moments of the neutron and
the proton \cite{Bartolini:2016dbk,Bartolini:2016jxq}: 
Despite the computation leading to the old chiral Lagrangian
cancellation issue ($\calDn=-\calDp$), the result was
obtained at leading order in few parameters of the theory, in
particular neglecting time derivatives, leaving open the possibility
of the appearance of a splitting in the magnitudes of the electric
dipole moments of the nucleons at the next-to-leading order or beyond. 

It is not a simple task to make an estimate of $\mathcal{D}_B$ since
it lies beyond the possibilities of the usual perturbative approach to
QCD, and even the lattice approach is tricky due to the presence of
the sign problem (as examples of a lattice estimate, see
Refs.~\cite{Shintani:2015vsx,Guo:2015tla,Alexandrou:2015spa,Shindler:2015aqa}):
throughout the years, many attempts with effective theories, such as
the chiral Lagrangian \cite{Crewther:1979pi} and the Skyrme model
\cite{Dixon:1990cq,Salomonson:1991ar} have achieved some good
estimates for the neutron.
Since the introduction of the $AdS/CFT$ duality by Maldacena in 1997
\cite{Maldacena:1997re}, it has been a major goal for theoretical
physicists to develop a holographic theory of QCD which could then be
used to explore its rich non-perturbative sector: the model which has
achieved the best degree of success so far is that of
Witten-Sakai-Sugimoto. 

The WSS model is based on a $D4$--$D8$ brane setup in type IIA string
theory.  
In the limit where a simple holographic dual description is given, the
model reduces to a $3+1$ dimensional large-$\Nc$ $SU(\Nc)$ gauge
theory with $\Nf$ massless quarks.  
Additionally, it also contains a tower of massive adjoint matter
fields whose mass scale is set by a dimensionful parameter denoted as
$M_{\KK}$ (which gives the scale of the glueballs as well). 
Despite this feature, at low energies, the model shares all the
expected features with QCD, like confinement, chiral symmetry breaking 
and so on.  
The WSS is the top-down holographic theory closest to QCD.
It incorporates automatically the whole tower of vector mesons and
exhibits complete vector dominance in the hadron electromagnetic form
factors.  It has very few parameters to fit. 
Flavor dynamics is encoded in the low-energy action for the gauge
field on the flavor branes, and the baryons of QCD are instantonic
configurations of that gauge theory
\cite{Hong:2007kx,Hata:2007mb,Hashimoto:2008zw,Bolognesi:2013nja}. 
Quantization of the degrees of freedom for an instantonic field of
charge one creates a quantum system of states, whose transformation
properties and quantum numbers are just right to interpret them as
nucleons.  
Nuclear physics at low energy is thus turned into a multi-instanton
problem in a curved five-dimensional background.

Just like baryons in the large-$\Nc$ limit can be seen as solitons of
the chiral Lagrangian, in the WSS model they are identified with
instantons of the holographic Lagrangian describing the mesonic sector
\cite{Hata:2007mb,Hong:2007kx}.

If quarks are massless, any $\theta$-dependence is washed out by a
chiral rotation of the quarks.  
A (small) mass term for the quarks can be introduced using a
prescription suggested in Refs.~\cite{Aharony:2008an,Hashimoto:2009hj}.

In this work we use the WSS model, supplemented with a finite quark
mass, to carry out a novel independent computation of $\calDD$ from first principles:
i.e.~the model of Witten-Sakai-Sugimoto adopts a top-down approach,
which provides us with valuable physical insights through the
calculations performed. It is, to our knowledge, the first holographic
attempt at performing this task.  

The paper is organized as follows. In Section~\ref{due} we will review
the main features of the nucleons in the WSS model, the inclusion of
the $\theta$ term and the electric dipole moment.  In
Section~\ref{tre} we perform the next-to-leading order analysis. In
Section~\ref{quattro} we use the newly found perturbations to compute
their contributions to the nucleon EDM showing that it is of isoscalar
nature. In Section~\ref{cinque} we relate the EDMs of the nucleons to
that of the deuteron.  We conclude in
Section~\ref{sei}. In Appendix~\ref{A} we provide the explicit form
of all the equations.   
In Appendix~\ref{B} we describe the numerical solution.

\section{Holographic QCD, nucleons and EDM}
\label{due}

\subsection{Background and effective action}
The starting point in the construction of the model is Witten's
confining background in type IIA supergravity: it is generated by a
stack of $\Nc$ coincident $D4$-branes, which encode color degrees of
freedom, making the theory holographically dual to $SU(\Nc)$ Yang-Mills.
The field content of the background includes the metric, the dilaton
and the Ramond-Ramond three-form $C_3$:  
\begin{equation}\label{metric}
\begin{split}
ds^2 &= \left(\frac{u}{R}\right)^{3/2} \left(\eta_{\mu\nu}dx^{\mu}dx^{\nu} + f(u)dx_4^2\right) + \left(\frac{R}{u}\right)^{3/2}\left( \frac{du^2}{f(u)} + u^2 d\Omega_4^2 \right), \\
e^{\phi}& = g_{\rm s} \left( \frac{u}{R} \right)^{3/4}, \quad\quad  F_4=dC_3= \frac{2\pi\Nc}{\Vol_4} \epsilon_4, \quad\quad f(u) = 1 - \frac{u_{\KK}^3}{u^3}.
\end{split}
\end{equation}
The $x_4$ and $u$ directions form a subspace with the shape of a
``cigar'', as can be seen from the fact that the geometry ends smoothly
at a finite value of the $u$ coordinate, viz.~$u=u_{\KK}$. 
The $x_4$ direction is compactified on an $S^1$ whose radius shrinks
to zero at $u=u_{\KK}$: absence of conical singularities fixes the
periodicity of the $x_4$ coordinate in terms of the radius of the
background $S^4$ (given by $R$ and fixed by the flux of $F_4$) and the
value of $u_{\KK}$ which is a free parameter. The relation is given by 
\beq
\delta x_4 = \frac{4\pi}{3}\frac{R^{3/2}}{u_{\KK}^{1/2}} \equiv \frac{2\pi}{M_{\KK}},
\eeq
where we have traded the free parameter $u_{\KK}$ for another one,
i.e.~the energy scale $M_{\KK}$ that defines the radius of $x_4$. It is
useful to work in units such that 
\beq
M_{\KK}= u_{\KK}=1,
\eeq
that is to say that we measure distances and energies in units of
$M_{\KK}^{-1}$ and $M_{\KK}$. Restoring the factors of $M_{\KK}$ at the
end of the computations will be easy using simple dimensional
analysis. 

The inclusion of flavor degrees of freedom is performed via the
addition of two stacks of $\Nf$ $D8/\overline{D8}$-branes in
the probe regime: we engineer them to be localized in the $x_4$
direction and antipodal on the $S^1$. This way the branes are found
to merge into a single stack at the cigar tip, realizing a holographic
version of chiral symmetry breaking. It is then useful to trade the
bulk coordinate $u$ with one that runs on the $D8$ world volume, call
it $z$, related by (in the antipodal setup) 
\begin{equation}
\left\{
\begin{aligned}
\quad  & u^3=u_{\KK}^3 + u_{\KK}r^2 \\
\quad & x_4= \frac{2 R^{3/2}}{3 u_{\KK}^{1/2}} \theta
\end{aligned} \right.
\quad\Rightarrow \quad\left\{
\begin{aligned}
\quad  & y=r \cos \theta \\
\quad & z=r \sin \theta
\end{aligned}
\right.
\end{equation}
The effective action at low energies is then given by the $D8$-branes
world-volume action in the curved background generated by the
$D4$-branes: after a trivial dimensional reduction on $S^4$, it amounts
to a Yang-Mills and Chern-Simons theory on a five-dimensional curved
space  
\bea\label{Sefftot}
S&=& S_{\YM} + S_{\CS},\nonumber \\
S_{\YM} &=&  -\kappa \Tr\int d^4 xdz \left[ \frac{1}{2}h(z)\mathcal{F}_{\mu \nu}^2 + k(z) \mathcal{F}_{\mu z}^2 \right],\\
S_{\CS} &=&  \frac{\Nc}{384 \pi^2} \epsilon_{\alpha_1 \alpha_2 \alpha_3 \alpha_4 \alpha_5} \int d^4 xdz \hA_{\alpha_1} \left[ 6\tr\left(F_{\alpha_2 \alpha_3}^a F_{\alpha_4 \alpha_5}^a \right)+2\tr\left(\hF_{\alpha_2 \alpha_3}\hF_{\alpha_4 \alpha_5}\right)\right],\nonumber
\eea
where $\kappa \equiv a \Nc \lambda$ with $a\equiv (216\pi^3)^{-1}$, and $k(z)= (1+z^2)$, $h(z)=k(z)^{-1/3}$.
In Eq.~\eqref{Sefftot} we introduced the $D8$ gauge field $\mA$, a
$U(\Nf)$ connection which we expand as 
\begin{equation}\label{fielddef}
\mathcal{A}= \hA \frac{\mathds{1}}{\sqrt{2\Nf}} + A^a T^a,
\end{equation}
where $T^a$ are the generators of $SU(\Nf)$ normalized as
$\tr(T^a T^b)=\frac{1}{2}\delta^{ab}$ (i.e.~$T^a = \frac{\tau^a}{2}$
in the $\Nf=2$ case). We adopt the following notation for space and
time indices: $\alpha$ labels all of the five directions of the
effective spacetime ($\alpha=0,\ldots,3,z$), Greek letters $\mu,\nu$
label the four-dimensional spacetime but not the bulk coordinate
($\mu,\nu=0,\ldots,3$), capital Latin letters label all spatial
directions ($M,N,\ldots=1,2,3,z$), while small Latin letters are
reserved for the three spatial directions that do not extend into the 
bulk ($i,j,\ldots=1,2,3$). 

\subsection{Baryons as holographic solitons}

Despite the model having mesons as fundamental degrees of freedom, it
can successfully describe baryons as a solitonic configuration with
a nontrivial instanton number. From a string theory point of view, this
would correspond to a $D4$-brane wrapped on $S^4$, with $\Nc$
fundamental strings connecting it to the color branes.

An approximate solution \cite{Hata:2007mb} is found by restricting the
analysis to a region near the cigar tip, where the warp factors $h(z)$
and $k(z)$ can be approximated by unity. This is a good approximation in
the large $\lambda$ limit since the baryon size is found to be of
order $\lambda^{-1/2}$. 
The static configuration is given by the $SU(2)$ BPST instanton in
flat space, with the addition of an electromagnetic potential in the
Abelian sector: 
\beq
A_M^{\rm cl} = -if(\xi) g\partial_M g^{-1}, \quad\quad
\hA_0 = \frac{\Nc}{8\pi^2\kappa}\frac{1}{\xi^2}\left[1-\frac{\rho^4}{(\xr)^2}\right],\quad\quad
A_0 = \hA_M = 0,
\eeq
with
\beq
f(\xi) = \frac{\xx}{\xr}, \quad\quad
g = \frac{\left(z-Z\right)-i\left(\vx-\vec{X}\right)\cdot\vt}{\xi},\quad\quad
\xi^2 =  \left(z-Z\right)^2+ |\vx -\vec{X}|^2.
\eeq
Note that $\rho$ and $Z$ are not real moduli of the soliton since they
have a potential 
\beq
U(\rho,Z) = 8\pi^2 \kappa \left(1+\frac{\rho^2}{6}+\frac{\Nc^2}{5(8\pi^2\kappa)^2\rho^2}+\frac{Z^2}{3}\right),
\eeq
which is minimized by the classical values 
\beq\label{classicalrhoz}
\rho^2_{\rm cl} =\frac{\Nc}{8\pi^2\kappa} \sqrt{\frac{6}{5}},\quad\quad
Z_{\rm cl}=0.
\eeq

Time dependence can be implemented in the moduli of the soliton:
$X^M(t)$ describes the position of the center of mass in
four-dimensional space, $\rho(t)$ is the instanton size, $y_I(t)$
describe the $SU(2)$ orientation: $y_I$ and $\rho$ are not
independent, they are actually related by $\sum y_I^2 = \rho^2$, so it
is useful to introduce $a_I \equiv y_I / \rho$. 
Other than promoting the moduli to be time-dependent quantities, a
transformation on the static gauge fields is also implemented: it
looks like a gauge transformation, but it is not since it does not act
on the $A_0$ field: 
\beq
A_M = VA_M^{\rm cl} V^{-1} - i V\de_M V^{-1}.
\eeq
This way the field strength transforms as:
\beq
F_{MN}&=&V F_{MN}^{\rm cl} V^{-1},\nn\\
F_{0M}&=&V \left(\dot{X}^{\alpha}\de_{\alpha}A_M^{\rm cl} - D_M^{\rm cl}\Phi\right)V^{-1},\label{F0M}
\eeq
with $\Phi$ given by
\beq\label{Phidef}
\Phi \equiv -iV^{-1}\dot{V}.
\eeq
To find a solution for $V(x,t)$ requires to find the function
$\Phi(x,t)$ and perform a path-ordered integration, but we will not
need this function, since $V(x,t)$ will only appear in our
computations in the form of $\Phi(x,t)$. 
A solution for the function $\Phi(x,t)$ is then found to be
\beq
\Phi(x,t) &=& -\dot{X}^N A_N^{\rm cl} + \chi^a (t) \Phi_a (x), \nn \\
\Phi_a &=& f(\xi) g\frac{\tau^a}{2}g^{-1},\nn \\
\chi^a &=& -i\tr(\A^{-1}\dot{\A}\tau^a),
\eeq
where the $SU(2)$ moduli only appear in the combination
$\A(t) = a_4 +ia_c \tau^c$.
The full time-dependent solution is given in singular gauge in
Ref.~\cite{Hashimoto:2008zw}: the motion of the center of mass is not
relevant for our computation, so we set $\dot{X}^M=\dot{\rho}=0$. Also
we will use the regular gauge, so our baryonic configuration reads 
\bea\label{fullbaryon}
A_M &=& -if(\xi) V\left(g\partial_M g^{-1}\right)V^{-1}-iV\de_M V^{-1}, \nn \\
A_0 &=& 0, \nn \\
\hA_i&=&-\frac{\Nc}{16\pi^2 \kappa }\frac{\rr}{(\xr)^2}\epsilon^{iab}\chi^a x^b,\nn\\
\hA_z&=&-\frac{\Nc}{16\pi^2 \kappa }\frac{\rr}{(\xr)^2}\vc \cdot \vx,\nn \\
\hA_0&=&\frac{\Nc}{8\pi^2 \kappa }\frac{1}{\xi^2}\left[1-\frac{\rho^4}{(\xr)^2}\right].
\eea
This configuration can be quantized in the moduli space approximation
to obtain the spectrum of baryons: the baryon states are labeled by
four quantum numbers $(l, I_3, n_{\rho}, n_z)$, to which, the third
component of the spin (labeled by $s$) and the three dimensional space
momentum $\vec{p}$, should be added for each baryon.
The spin and isospin operators are constructed in terms of the $SU(2)$
moduli $y_I$ as: 
\bea
\label{defI}
I_a &=& \frac{i}{2} \left(y_4 \frac{\de}{\de y_a} - y_a \frac{\de}{\de y_4} -\epsilon^{abc} y_b \frac{\de}{\de y_c}\right),\\
\label{defJ}
J_a &=&\frac{i}{2} \left(-y_4 \frac{\de}{\de y_a} + y_a \frac{\de}{\de y_4} -\epsilon^{abc} y_b \frac{\de}{\de y_c}\right),
\eea
from which it follows that $I^2 = J^2$ so only states with $I=J=l/2$
enter the spectrum. 
The moduli $y_I$ are related to their canonical momenta by
\beq\label{ymomenta}
\Pi_I = -i\frac{\de}{\de y_I} = 16\pi^2\kappa \dot{y}_I.
\eeq
Using the definition of $a_I$, and Eqs.~\eqref{defI}, \eqref{defJ} and
\eqref{ymomenta}, we can write down the following relations: 
\begin{equation}
I_k = -i 4\pi^2\kappa\rho^2\tr\left( \A\dot{\A}^{-1}\tau^k \right) \quad \Rightarrow\quad \A\dot{\A}^{-1} = \frac{i}{8\pi^2\kappa\rho^2} \left(\vec{I}\cdot \vec{\tau}\right),
\end{equation}
\begin{equation}\label{Jchi}
J_k = -i 4\pi^2\kappa\rho^2 \tr\left(\A^{-1}\dot{\A}\tau^k\right)=   4\pi^2\kappa\rho^2 \chi^k.
\end{equation}
Finally, we recall that another useful gauge choice is the singular one: we will use it later in the development of the set of equations to be solved.
It is reached from the regular gauge by a transformation
\beq \label{singgauge}
A_{\alpha} \rightarrow GA_{\alpha}G^{-1} - i G\de_{\alpha}G^{-1},
\eeq
with $G= \A(t)gV^{-1}$. In this gauge the $SU(2)$ moduli $\A$ appear
explicitly in the field configuration rather than being ``hidden'' in
the asymptotics of the function $V$, making it easier to use all the
machinery developed in the context of other solitonic models of
baryons.

We will often exploit the relation $g\xt g^{-1}=g^{-1}\xt g = \xt$
since this quantity will appear often after gauge transformations of
both the source terms introduced by finite quark-mass deformation, and
the perturbations it induces. The explicit form of the fields in this
gauge can be computed from Eqs.~\eqref{fullbaryon} and
\eqref{singgauge}, but we will not need it throughout this article.

\subsection{Quark masses}

The presence of the $D8$-branes alone accounts for the inclusion of
massless quarks in the model: We know from QCD that in this setup the
chiral anomaly eliminates the dependence on $\theta$ from physical
observables, thus making every CP violating quantity vanish, such as
intrinsic electric dipole moments.
To include $\theta$ dependence in the model, we need to account for
nonvanishing bare masses for each flavor. This deformation of the
$D4$--$D8$ setup was explored in Ref.~\cite{Aharony:2008an}: An open Wilson
line operator on the field theory side is dual to a fundamental string
worldsheet whose boundary is given by said Wilson line. 

In the Sakai-Sugimoto model, the Wilson line stretches along the $x_4$
direction between the two stacks of $D8$-branes, i.e.~the string
worldsheet extending in the cigar subspace.
This is realized by adding the following term to the action
\beq\label{SAK}
S_{\rm AK} = c\int d^4x
\Tr\mathcal{P}\left[M_{2\times2}e^{-i\int_{-\infty}^{+\infty}dz\mA_z}+{\rm h.c}\right]; \quad\quad
c=\frac{\lambda^{3/2} }{3^{9/2}\pi^3}.
\eeq
We will work in the mass-degenerate scenario, since we are not
interested in the effects of explicit isospin breaking, and hence we
can identify 
\beq
M_{2\times2} = m\mathds{1}_{2\times2}.
\eeq
In the antipodal setup of the flavor branes, this is the only effect
we need to take into account: of course the string tension would deform
the shape of their embedding in the cigar space, but in this
particularly symmetric setup, the contributions from strings on both
sides of the $x^4$ circle are equal and thus cancel out.

\subsection{Holographic \texorpdfstring{$\theta$}{theta}-term}

This holographic model can successfully account for the presence of a
QCD $\theta$ term. This can be seen by looking at the action for the
color $D4$-branes: it includes a coupling to the Ramond-Ramond
$1$-form, $C_1$, given by 
\beq
S^{\rm WZ}_{D4-C_1}= \frac{(2\pi \alpha')^2}{2!}\mu_4 \tr\int_{\mathcal{M}_4\times S^1} C_1\wedge G\wedge G.
\eeq
If we take the $G_{\mu\nu}$ components of the $D4$ gauge field to
correspond to the QCD gluonic field strength, then the $x_4$ component
of $C_1$, after integration, plays the role of a $\theta$ angle: 
\beq
\int_{S^1_{x_4}} C_1 = \theta +2\pi k.
\eeq
The reproduction of the shift of $\theta$ under an axial chiral
transformation is also included through a nontrivial mechanism of
anomaly inflow: in the presence of the flavor branes, the $C_7$
Ramond-Ramond form action includes, other than a kinetic term, a
coupling to the flavor gauge field $\hA$:
\beq
S_{C_7} = -\frac{1}{4\pi} (2\pi l_{\rm s})^6 \int dC_7 \wedge \star dC_7
+\frac{1}{2\pi} \int C_7 \wedge \tr\mathcal{F}\wedge \omega_y,
\eeq
where $\omega_y$ is a form that describes the distribution of the
branes in the $y$ direction of the cigar (i.e.~in our setup it is
simply $\omega_y= \delta(y)dy$). 
The coupling of $C_7$ to the trace part of the flavor gauge field
translates into an anomalous Bianchi identity for the field strength
$\tilde{F}_2$ related to $F_8=dC_7$ by Hodge duality 
\beq\label{F2bianchi}
d\tilde{F}_2 = \tr\mathcal{F}\wedge \omega_y.
\eeq
This equation can be solved by giving up the condition that
$\tilde{F}_2=dC_1$ (this is why we used the tilde notation: we would
call $F_2=dC_1$, while $\tilde{F}_2$ corresponds to the solution of 
Eq.~\eqref{F2bianchi}), so that $\tilde{F}_2$ reads 
\beq
\tilde{F}_2 = dC_1 + \tr\mathcal{A}\wedge\omega_y
= dC_1 + \sqrt{\frac{\Nf}{2}}\hA\wedge \delta(y)dy.
\eeq
This formula implies that the presence of $D8$-branes makes the form
$C_1$ a non-gauge invariant quantity: only $\tilde{F}_2$ is gauge
invariant. A gauge transformation along the $z$ direction reduces on
the UV boundary to an axial transformation, hence reproducing the
shift of the $\theta$ angle. 
If the fermions are massive, we expect the shifted $\theta$ to appear
as a phase in the mass matrix of the quarks: it is easy to see that
the action \eqref{SAK} reproduces exactly this feature when the
corresponding gauge transformation is performed on $\hA_z$.

\subsection{Nucleon EDM at leading order}

Here we briefly review the results of Refs.~\cite{Bartolini:2016dbk,Bartolini:2016jxq},
i.e.~the leading order EDM of the nucleons, which will be the 
starting point from which to build and expand in order to obtain an
estimate for the deuteron EDM. From now on, we set $\Nf=2$. 

The first thing to notice is that the $\hA_z$ vacuum in presence of
$\theta$ term is nontrivial: adopting a pure gauge Ansatz for it, such
as $\hA^{\rm vac} = f(z)dz$, the supergravity action for $\tilde{F}_2$
imposes the following condition through the equation of motion
(integrated over $z$):
\beq\label{thetavac}
-\frac{1}{2}\int dz \, \hA_z^{\rm vac} = \frac{\theta}{2}.
\eeq
From now on we define:
\beq
\tilde{\varphi}(r) \equiv -\frac{1}{2}\int dz \left(\hA^{\rm vac}_z+\hA_z\right) = \frac{\theta}{2} + \varphi(r).
\eeq
The function $\tilde{\varphi}$ will enter the equations of motion
through the mass term \eqref{SAK}, thus generating $\theta$-dependent
perturbations in the baryon configuration of the fields. We use the
unperturbed baryon configuration to evaluate this term (i.e.~we
neglect terms of order $m^2$): This term will be a source for the
first-order mass perturbation of the baryon. 

It is possible to identify the pion field with:
\beq
\pi^a(x) = -\frac{f_\pi}{2}\int^{+\infty}_{-\infty}dz \, A_z^a.
\eeq
So we can actually identify the holonomy appearing in Eq.~\eqref{SAK} with 
\begin{equation}
e^{-i\int_{-\infty}^{+\infty}dz \mathcal{A}_z} \equiv e^{i\left(\frac{\theta}{2}+\varphi\right)}U,
\end{equation}
where we have made use of Eq.~\eqref{thetavac}.

Plugging in the baryon configuration (in singular gauge, which we will
use in the rest of this section) with full time dependence, we can
write the pion matrix $U$ as 
\begin{equation}
U = \exp\left[-i\pi\A\xt\A^{-1}\left(1-\a\right)\right]=-\cos\alpha -i\sin \alpha \frac{x^a}{r}\A \tau^a \A^{-1};\quad\quad \alpha \equiv \frac{\pi}{\sqrt{1+\frac{\rr}{r^2}}}.
\end{equation}
The equations of motion, in singular gauge, for the $\mA_z$ fields read:
\bea
-\kappa k(z) \partial_{\mu}\hF^{z\mu}+\left(\CS\right)&=& 2 cm(\cos\alpha+1)\sin\tilde{\varphi}, \label{abelz}\\
-\kappa k(z)\left[D_{\nu} F^{z\nu}\right]^a+ \left(\CS\right)&=&  cm\sin\alpha \cos\tilde{\varphi } \frac{x^k}{r}\tr\left(\A\tau^k\A^{-1}\tau^a\right).
\eea
In these equations, we neglected the Chern-Simons term, regarding each
coordinate as being of the order $x_M\sim \lambda^{-1/2}$ and
correspondingly each field $A_M \sim \lambda^{1/2}$. 

We now extract the $\theta$ dependence by expanding
$\sin\tilde{\varphi}$ and $\cos\tilde{\varphi}$ to first order in
$\theta$ obtaining the set  
\begin{equation}
\begin{split}
-\kappa k(z) \partial_{\mu}\hF^{z\mu}+\left(\text{CS}\right)&=  cm\theta(\cos\alpha+1)\cos\varphi,\\
-\kappa k(z)D_{\nu} F^{z\nu}+ \left(\text{CS}\right)&= -\frac{cm\theta}{2}\sin\alpha \sin \varphi \frac{x^k}{r}\A\tau^k\A^{-1}.
\end{split}
\end{equation}
We employ a perturbative approach, expanding every field as
$\mA =\mA^{\rm bar} + \delta\mA$ where $\delta \mA$ is intended to be
linear in $m\theta$ and $\mA^{\rm bar}$ is the unperturbed baryon
configuration. 
Let us now neglect time derivatives of the moduli for the moment: if
we do so, we can approximate $\cos\varphi \sim 1$ and
$\sin\varphi\sim 0$, so that only the Abelian field $\hA_z$ will have
a source term which is linear in $\theta$. A solution to the equations
of motion (consistent with the ones for $\hA_i$ and $A_i$) in this
approximation is given by 
\beq \label{deltaAz}
\delta \hA_z &=& \frac{cm\theta}{\kappa}\frac{ u(r)}{k(z)},\\
\delta A_M &=& 0,
\eeq
with $u(r)$ defined by
\beq
\nabla^2 u(r) = \cos\alpha + 1.
\eeq
This equation can be solved via the Green's function:
\beq
u_G(r,r') = \left\{\begin{array}{ll}
	-r', & r<r',\\
	-r'\left(\frac{r'}{r}\right), & r>r'.
	\end{array}\right.
\eeq
Then the solution is given by:
\beq\label{defu}
u(r) = \int_{0}^{+\infty}dr'\, u_G(r,r')\left(1+\cos \frac{\pi}{\sqrt{1+\rho^2/{r'}^2}}\right).
\eeq
However, we did not analyze every equation of motion yet: we still
need to solve the one for $A_0$. For this equation the Chern-Simons
term is not subleading in $\lambda$, and it contains the Abelian field
strength $\hF_{zk}$: The newly found perturbation \eqref{deltaAz} will
then produce a source for $\delta A_0$ when inserted in this term.  
The full equation reads:
\beq
-\kappa\left(h(z)\delta\left[D_iF^{0i}\right] +\delta \left[D_z\left(k(z)D_zF^{0z}\right)\right]\right)^a +\frac{\Nc}{32\pi^2}\epsilon^{ijk}F^a_{ij}\delta \hF_{zk}=0.
\eeq

Employing the Ansatz
\beq\label{deltaA0sing}
\delta A^0 =27 \pi \frac{cm\theta}{\lambda\kappa} \A W (\vec{x}\cdot  \vec{\tau})\A^{-1},
\eeq
we find the following equation for the function $W(r,z)$ to be solved numerically
\begin{align}
h(z)\left(W'' + \frac{4}{r}W' + \frac{8\rr}{(\xr)^2}W \right)  +  \partial_z \left(k(z)\de_zW\right) = \frac{\rr}{(\xr)^2} \frac{1}{r}\frac{u'}{k(z)}.
\label{defW}
\end{align}

\begin{figure}[!t]
	\centering
	\includegraphics[width=0.5\linewidth]{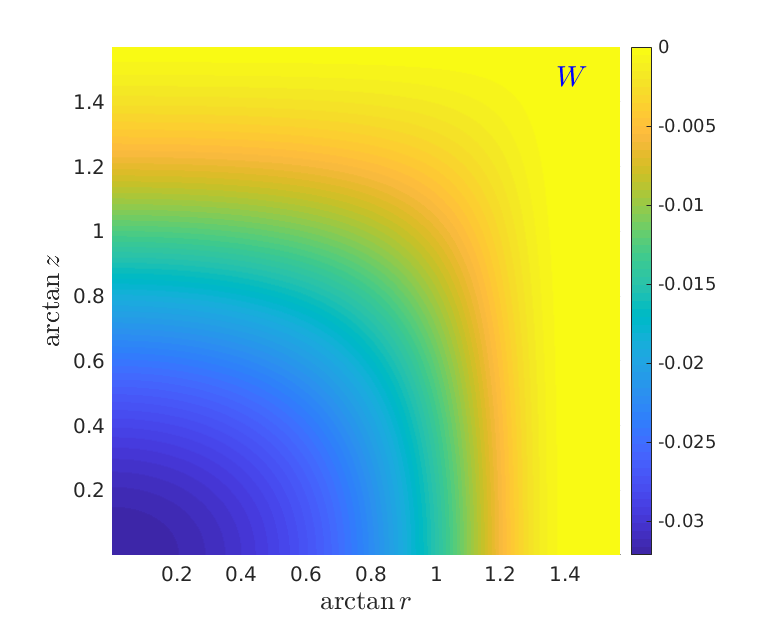}
	\caption{\small The function $W(r,z)$ that solves Eq.~\eqref{defW}
          for the semiclassical value of the size $\rho=\rho_{cl}$. }
        \label{Wplot}
\end{figure}

It is precisely the function $W(r,z)$ that will produce the
leading-order term in the EDM of the nucleons. A numerical solution is shown in Fig.~\ref{Wplot}. The electromagnetic
holographic current is given by 
\beq \label{Jem}
J^{\mu}_{\rm em} = \tr \left(J^{\mu}_V \tau^3\right) + \frac{1}{\Nc}\widehat{J}^{\mu}_V,
\eeq
where $\mathcal{J}^{\mu}_V$ is defined as
\beq\label{JV}
\mathcal{J}^{\mu}_V = -\kappa \left[k(z)\mF^{\mu z}\right]^{+\infty}_{-\infty}.
\eeq
Of this current we are interested in the component $J^0_{\rm em}$,
since we want to compute the EDM of nucleons, defined by
\beq
\calDN^i =  e\int d^3x\, x^i \langle {\rm N}| J^0_{em}
|{\rm N} \rangle = \calDN \langle s|\sigma^i|s \rangle.
\eeq
The EDM of the nucleon will consist of two terms with different symmetry properties under isospin transformations: we employ the following notation for these isovectorial and isoscalar parts:
\beq\label{notationEDM}
\calDN = d_N + \Delta_N \quad;\quad d_n=-d_p\quad;\quad \Delta_n= \Delta_p 
\eeq
We can immediately see that the Abelian part of Eq.~\eqref{Jem}
vanishes with the approximation employed, while the part constructed
with the non-Abelian fields will contribute, having precisely a dipole
structure \eqref{deltaA0sing}. From this observation alone, we can
already predict that the EDM $\calDN$ will be proportional to the
third component of the isospin operator $I_3$, and hence will be of
equal magnitude and opposite sign for proton and neutron, thus only $d_N$ is nonvanishing at this order. 

The computation confirms this, yielding the semiclassical ($Z=0$,
$\rho=\rho_{\rm s.c.}$) results presented in \cite{Bartolini:2016dbk}:
\beq
d_n = -d_p = 0.78\times 10^{-16}\theta\text{ $e\cdot$cm}.
\eeq
Effects of the nucleon wave function have been included in the results of \cite{Bartolini:2016jxq} and turn out to
be quantitatively important, but are not relevant for the purpose of
this article since they do not change the isovectorial nature of
$\calDN$ at this level of approximation.

\section{Perturbing the baryon at NLO}
\label{tre}

We will now take the perturbative approach to the next-to-leading
order. 
The values of the parameters $\theta$ and $m/M_{\KK}$ will
remain small, also in the phenomenologically relevant portion of
the parameter space, so we will still keep terms which are first order
with respect to them. On the other hand, higher orders in
$\lambda^{-1}, \Nc^{-1}$ will provide relevant corrections, in
particular the leading contribution to the splitting
of the magnitude of EDMs of nucleons. 

For a field to give a nonvanishing EDM, it must be odd in the $\vx$
coordinate: Since the holographic electromagnetic current is built
from $\mF_{z0}$, we are looking for perturbations in any field $\mA$
that can result in perturbations $\delta \mA_{z}, \delta \mA_0$ which
are odd in $\vx$. Since those fields are scalars under
three-dimensional spatial rotations, the odd powers of $\vx$ should come
in scalar products (or combined with the antisymmetric tensor
$\epsilon^{ijk}$) with other vectors: natural guesses are the angular
velocity $\vc$, the isospin $\vec{I}$ and the $SU(2)$ generators
$\vec{\tau}$. 

As shown by the results for the leading-order contribution to the
NEDM, a $\delta A_{0,z}\propto \A\vx\cdot\vt \A^{-1}$ would not
produce any splitting in the EDM magnitudes.  
More generally, it can be stated that the $SU(2)$ part alone of the
current $J_{\mu}^{a=3}$ cannot produce a splitting of the EDMs due to its
symmetry properties: Once evaluated on isospin eigenstates (i.e.~the
nucleons) it is bound to give results proportional to $I_3$, hence
producing EDMs of equal magnitude (and opposite sign).
The Abelian part of the current $\widehat{J}_\mu$ instead is
an isoscalar: It acts blindly on nucleon states, so that also its action
alone would produce EDMs of equal magnitude (and equal sign). 
When both terms are present, their combination is not isovectorial nor
isoscalar, hence the EDMs will be split in magnitude. 

Since the leading result for the nucleon EDM is given by the $SU(2)$
current, we now look for the leading $\theta$-dependent contribution
to $\widehat{J}_0$.
The only possible spatial vectors that can appear in $\widehat{J}_0$
are $\vc$ and $\vx$, hence we will look for perturbations in all the
fields that can lead to a dipole structure  
\beq
\widehat{J}_0 \propto \mathcal{M}(r,z) (\vec{x}\cdot\vec{\chi}).
\eeq
It is now clear in what sense we need to move to the next-to-leading
order: since $\vec{\chi}$ is first order in time derivatives (which
are to be regarded as $\Nc^{-1}$), we will now include such terms in
the equations of motion and neglect higher-order terms. This means
that we cannot drop time derivatives in the Yang-Mills part anymore,
and we cannot approximate $\sin\varphi\sim0$, but instead we need to
include $\sin\varphi\sim\varphi$. Since we are stopping at the linear
order in time derivatives, we can still approximate
$\cos\varphi\sim 1$. 







With this in mind, we can move to look at the equations of motion and
seek for terms that could work as sources for the perturbations of
order $m\theta$.

\subsection{Relevant equations}

We begin by recalling the equations in singular gauge, starting with
the ones with explicit source terms coming from the Aharony-Kutasov
action (i.e.~the ones for $\mA_z$). Up to first order in time
derivatives and in the limit of small $\varphi$, they read 
\bea
-\kappa k(z) \partial_{\mu}\hF^{z\mu}+\left(\CS\right)&=&  cm\theta(\cos\alpha+1), \label{eomhz} \nn \\
-\kappa k(z)D_{\nu} F^{z\nu}+ \left(\CS\right)&=&
-\frac{cm\theta}{2}\sin(\alpha) \varphi\, \A\xt\A^{-1}, \label{eomz} 
\eea
with
\beq\label{varphi}
\varphi = -\frac{1}{2}\int^{+\infty}_{-\infty} dz \,\hA_z  -\frac{\theta}{2}= \frac{\Nc}{64\pi \kappa} \frac{\rr}{(\rr+r^2)^{3/2}}r\xc.
\eeq
The other equations we are interested in are
\bea\label{eoms}
&&-\kappa \left[h(z)D_{\nu} F^{i\nu} + D_z\left(k(z)F^{iz}\right)\right]+\left(\CS\right) =0, \label{eomi}  \\
&&-\kappa \left[h(z) \partial_{\nu}\hF^{0\nu} + \partial_z\left(k(z)\hF^{0z}\right) \right]-\frac{\Nc}{32\pi^2}\epsilon^{ijk}\left(F^a_{ij}F^a_{kz} + \hF_{ij}\hF_{kz}\right)=0.\label{eomh0}
\eea
The current $\widehat{J}_0$ we are interested in, is built from the
field strength $\hF_{z0}$: still, the $\delta\hA_z$ field is
suppressed with a time derivative, and also cannot acquire both 
a factor $\vc$ and $\theta$ as can be argued from Eq.~\eqref{eomhz}. The
only perturbation that will directly enter the current is then
$\delta\hA_0$, but we will keep the leading order solution for
$\delta\hA_z$ given by Eq.~\eqref{deltaAz}. 

As in Ref.~\cite{Bartolini:2016jxq}, the Chern-Simons term will act as
a source for this perturbation: the Abelian part of the Chern-Simons term
in Eq.~\eqref{eomh0} reads 
\beq\label{abelianCS}
-\frac{\Nc}{32\pi^2}\epsilon^{ijk}\hF_{ij}\delta\hF_{kz} =\frac{\Nc}{16\pi^2}\frac{\rr}{(\xr)^2}\frac{1}{r}\de_r\delta \hA_z\left(\vec{x}\cdot\vc\right).
\eeq
Hence it is linear both in $\theta$ and $\vc$ as desired.

However, at the same order, new sources may appear from the non-Abelian
fields in the same Chern-Simons term: since the unperturbed field
strength $F_{MN}$ does not contain neither $\vc$ nor $\theta$, the
perturbed $\delta A_M$ can only contribute if they are of order
$\theta \vc$ themselves. In the next section, we show how
Eqs.~\eqref{eomi} and \eqref{eomz} precisely contain sources of
that order and must then be solved before moving to perturb
Eq.~\eqref{eomh0}.

\subsection{Sources for \texorpdfstring{$\delta A_M$}{(delta A)\_M}}

The possible source terms come from two parts of the equations: the
perturbed Yang-Mills terms containing $\delta A_0$, and the
Aharony-Kutasov term (since the function $\varphi$ contains
$\vec{\chi}$). 
We will compute the Yang-Mills part in regular gauge for the sake of
simplicity and for avoiding possible singularities in the numerical
integration that will follow. 

The perturbation $\delta A_0$ was obtained in
Refs.~\cite{Bartolini:2016dbk,Bartolini:2016jxq} in singular gauge, but
it is simple to bring it back to the regular one, since the
transformation acts on $\delta A_0$ as: 
\beq
\delta A_0^{\rm (reg)} = Vg\,\A^{-1} \delta A_0^{\rm (sing)}\A\, g^{-1} V^{-1} = W(r,z) V \xt V^{-1}.
\eeq
The field $\delta A^0$ \eqref{deltaA0sing} is already of the order of
$\theta$, and appears in Eqs.~\eqref{eomz} and \eqref{eomi} with time
derivatives, that will act on the functions $V,V^{-1}$ to generate
$\Phi(x,z)$. 


We will not follow the usual approach of solving first the static
equations and then implementing time dependence modifying the static
solution: We already know that we want to keep time derivatives up to
first order, so we use the following Ansatz for the time dependence of
the perturbed non-Abelian fields 
\beq\label{factorV}
\delta A(x,z,t) \equiv V \delta \widetilde{A}(x,z,\vc) V^{-1}.
\eeq
The field $\delta A_0$ also shares this very same form if we consider
$\delta \widetilde{A}_0 = W(\vec{x}\cdot\vec{\chi})$.

The unperturbed fields are instead of the form
\beq
A(x,z,t) \equiv V A^{\rm cl}_M V^{-1} - iV\de_M V^{-1}.
\eeq


With these choices, the functions $V,V^{-1}$ can be factorized out respectively on the left and the right of the full perturbed Yang-Mills term as follows:
\beq\label{pertM}
&&-\kappa k(z)V\left\{D_j^{\rm cl}\delta\widetilde{F}^{zj}+i\left[\delta \widetilde{A}_j,F_{\rm cl}^{zj}\right]+i\left[\Phi,\de_z\delta\widetilde{A}^0\right]\right. \nonumber\\
&&\label{perz} \left.\quad -\left[\Phi,\left[A_z^{\rm cl},\delta \widetilde{A}^0\right]\right]+i\left[\delta \widetilde{A}_0,F_{\rm cl}^{z0}\right]\right\}V^{-1}= (\text{AK term})^{\rm reg},\\
&&-\kappa h(z)V\left\{D_j^{\rm cl}\delta \widetilde{F}^{ij}+i\left[\delta \widetilde{A}_j,F_{\rm cl}^{ij}\right]\right\}V^{-1}\nonumber\\
&&\quad-\kappa V\left\{k(z)D_z^{\rm cl}\delta\widetilde{F}^{iz}+2z\delta \widetilde{F}^{iz}+k(z)i\left[\delta \widetilde{A}_z,F^{iz}_{\rm cl}\right]\right\}V^{-1}\nonumber\\
&&\quad-\kappa h(z)V\left\{i\left[\Phi,\de_i\delta \widetilde{A}^0\right]-\left[\Phi,\left[A_i^{\rm cl},\delta \widetilde{A}^0\right]\right]+i\left[\delta \widetilde{A}_0, F_{\rm cl}^{i0} \right]\right\} V^{-1}=0, \label{peri}
\eeq
where we have neglected second-order time derivatives and made use of
Eqs.~\eqref{F0M} and \eqref{Phidef}. We do not need the explicit expression
of the Aharony-Kutasov term in this gauge. It is evident that the last
row of every equation is now a source term for the new perturbations
$\delta \widetilde{A}_M$: However, cast this way, the equations are
hard to solve, since we would need the full knowledge of the function
$V(x,z,t)$. 
To overcome this problem we now transform the gauge back to 
singular gauge: the Yang-Mills term transforms covariantly, so it is
simply obtained by the substitution $V \rightarrow \A\, g^{-1}$.
In singular gauge, we already computed the Aharony-Kutasov term, so we
can restore its explicit form. 
Putting all the pieces together, we finally obtain the following set of
equations: 
\beq
&&\A\, g^{-1}\left\{D_j^{\rm cl}\delta\widetilde{F}^{zj}+i\left[\delta \widetilde{A}_j,F_{\rm cl}^{zj}\right]\right. \nonumber\\
&&\quad \left.+i\left[\Phi,\de_z\delta\widetilde{A}^0\right]-\left[\Phi,\left[A_z^{\rm cl},\delta \widetilde{A}^0\right]\right]+i\left[\delta \widetilde{A}_0,F_{\rm cl}^{z0}\right]\right\}g\,\A^{-1}\nonumber\\
&&\quad -\A \left[\frac{\Nc cm\theta}{128\pi \kappa^2} \frac{\rr}{(\rr+r^2)^{3/2}}\frac{r}{k(z)}\sin \alpha \xc\xt\right] \A^{-1}=0,\label{perzintermediate}\\ 
&& h(z)\,\A\, g^{-1}\left\{D_j^{\rm cl}\delta \widetilde{F}^{ij}+i\left[\delta \widetilde{A}_j,F_{\rm cl}^{ij}\right]\right\}g\,\A^{-1}\nonumber\\
&&\quad +\A\, g^{-1}\left\{k(z)D_z^{\rm cl}\delta\widetilde{F}^{iz}+2z\delta \widetilde{F}^{iz}+k(z)i\left[\delta \widetilde{A}_z,F^{iz}_{\rm cl}\right]\right\}g\,\A^{-1}\nonumber\\
&&\quad +h(z)\,\A\, g^{-1}\left\{i\left[\Phi,\de_i\delta \widetilde{A}^0\right]-\left[\Phi,\left[A_i^{\rm cl},\delta \widetilde{A}^0\right]\right]+i\left[\delta \widetilde{A}_0, F_{\rm cl}^{i0} \right]\right\}g\,\A^{-1}=0. 	\label{periintermediate}
\eeq
As can be seen, the last two rows of Eq.~\eqref{perzintermediate} and the last
row of Eq.~\eqref{periintermediate} are the source terms we were looking for: we
now only need to factorize away the $\A,\A^{-1}$ on each side of the
equations, and exploit the fact that $g^{-1}\xt g = g\xt g^{-1} =
\xt$, to finally obtain our set of equations to solve 
\beq\label{pertfM}
&&\left\{D_j^{\rm cl}\delta\widetilde{F}^{zj}+i\left[\delta \widetilde{A}_j,F_{\rm cl}^{zj}\right]\right. \nonumber\\
&&\quad \left.+i\left[\Phi,\de_z\delta\widetilde{A}^0\right]-\left[\Phi,\left[A_z^{\rm cl},\delta \widetilde{A}^0\right]\right]+i\left[\delta \widetilde{A}_0,F_{\rm cl}^{z0}\right]\right\}\nonumber\\
&&\label{perzfinal}\quad -\frac{\Nc cm\theta}{128\pi \kappa^2} \frac{\rr}{(\rr+r^2)^{3/2}}\frac{r}{k(z)}\sin \alpha \xc\xt =0,\\
&& h(z)\left\{D_j^{\rm cl}\delta \widetilde{F}^{ij}+i\left[\delta \widetilde{A}_j,F_{\rm cl}^{ij}\right]\right\}\nonumber\\
&& \quad+\left\{k(z)D_z^{\rm cl}\delta\widetilde{F}^{iz}+2z\delta \widetilde{F}^{iz}+k(z)i\left[\delta \widetilde{A}_z,F^{iz}_{\rm cl}\right]\right\}\nonumber\\
&& \quad+h(z)\left\{i\left[\Phi,\de_i\delta \widetilde{A}^0\right]-\left[\Phi,\left[A_i^{\rm cl},\delta \widetilde{A}^0\right]\right]+i\left[\delta \widetilde{A}_0, F_{\rm cl}^{i0} \right]\right\}=0. \label{perifinal}
\eeq

\subsection{The Ansatz}

We now want to solve the Eqs.~\eqref{perzfinal} and \eqref{perifinal}:
doing so is not an easy task since they are actually twelve coupled
differential equations in four variables. Luckily enough, symmetry can
be exploited to construct suitable Ans\"atze for the fields
$\delta\widetilde{A}_M$: First of all, we note that three-dimensional radial
symmetry of each field is only broken by the presence of the vectors
$\vec{\chi}$ and $\vec{\tau}$. This means we can construct every
structure that combines $\vec{\chi},\vec{\tau}$ and $\vec{x}$, and
multiply each one of them by a function of $(r,z)$. 
\beq \label{pertAz}
\delta \widetilde{A}_z \equiv K\left\{\beta(r,z) \xc \xt + \gamma(r,z) \left(\vc \cdot \vt\right) +\delta(r,z) \abc\right\},
\eeq
\bea \label{pertAi}
\delta \widetilde{A}_i \equiv K&\Big\{&B(r,z)\;\chi^i\xt  \nonumber\\
&+& C(r,z)\xc \tau^i \nonumber\\
&+& D(r,z)\;\hat{r}^i \ct\nonumber\\
&+& E(r,z)\;\eiab \chi^a\tau^b\nonumber\\
&+& F(r,z)\;\hat{r}^i\xc\xt\nonumber\\
&+& G(r,z)\;\hat{r}^i\abc\nonumber\\
&+& H(r,z)\;\eiab \chi^a\hat{r}^b\xt\nonumber\\
&+&I(r,z)\xc\eiab \hat{r}^a \tau^b\Big\}.
\eea
We choose to use unit vectors $\hat{r}$ instead of $\vec{x}$. With
this choice it will be easier to impose regularity of the fields at
$r=0$, which will translate into simple Neumann conditions for the
radial functions (exploiting $\de_r \hat{r}=0$), and also every
function will now have the same length dimension, regardless of how
many coordinate vectors enter the respective group structure.

The complete set of eleven equations (with the corresponding boundary
conditions) originating from this Ansatz plugged into
Eqs.~\eqref{perzfinal} and \eqref{perifinal} is given in Appendix \ref{A}. 
Since one of the fields that act as a source in this case is given by
Eq.~\eqref{deltaA0sing}, we also choose the overall constant
(factorized away in the equations in Appendix \ref{A}) to be 
\beq
K \equiv  \frac{27 \pi cm \theta}{\lambda \kappa} = \frac{\Nc cm\theta}{8\pi^2\kappa^2}.
\eeq

The Ansatz for the field $\delta \hA^0 $ is easier since now there is
no group structure: the only possibility is 
\beq\label{perthA0}
\delta \hA^0 \equiv \Upsilon \mathcal{M}(r,z)\xc,
\eeq
and since the perturbed fields $\delta\widetilde{A}_M$ appear as
sources in Eq.~\eqref{eomh0} via the Chern-Simons term, we choose the
overall constant $\Upsilon$ to be 
\beq
\Upsilon \equiv \frac{\Nc K}{32\pi^2\kappa} = \frac{\Nc^2 cm\theta}{256\pi^4 \kappa^3}.
\eeq
With all these choices, the resulting equation for $\mathcal{M}$
obtained by plugging Eqs.~\eqref{pertAi}, \eqref{pertAz}, \eqref{deltaAz}
and \eqref{perthA0} into Eq.~\eqref{eomh0} reads 
\beq
&&-h(z)\left(\mM'' + \frac{2}{r}\mM'-\frac{2}{r^2}\mM\right) -2z\dot{\mM}-k(z)\ddot{\mM}\nonumber\\
&&\quad+\frac{16\rr}{(\xr)^2}\left(2E'+\frac{2}{r}G-2H+2I'+\frac{4}{r}I+\beta'\right.\nonumber\\
&&\phantom{\quad+\frac{16\rr}{(\xr)^2}\bigg(}\left.+\frac{2}{r}\beta+\gamma'-\dot{B}-3\dot{C}-\dot{D}-\dot{F}-\frac{1}{8}\frac{u'}{k(z)}\right)\nonumber\\
&&\quad+\frac{64\rr}{(\xr)^3}\left(zB+3zC+zD-2rE-2rI-r\beta-r\gamma\right) = 0. \label{eomM}
\eeq
Consistency requires that all the perturbations we turned on, do not
change the baryonic number of the soliton solution. This is trivially
guaranteed by the dipole structure of the perturbation: the baryon number
density is given by the isoscalar charge density as 
\beq\label{baryondensity}
J^0_B \equiv -\frac{2}{\Nc}\kappa \left[k(z)\hF^{0z}\right]^{z=+\infty}_{z=-\infty}
\eeq
so its perturbation amounts to:
\beq
\delta J^0_B = \frac{2}{\Nc}\kappa \left[k(z)\de_z\delta \hA^{0}\right]^{z=+\infty}_{z=-\infty} =   \frac{2}{\Nc}\kappa \Upsilon\left[k(z)\de_z\mathcal{M}\xc\right]^{z=+\infty}_{z=-\infty}
\eeq
which is odd in $\vec{x}$ and thus vanishes upon integration over the solid angle.

\section{Neutron-proton EDM splitting}
\label{quattro}

We now move to compute the splitting in the EDM magnitude of the
nucleons: we recall the definition of the electric dipole moment for a
baryon: 
\beq
\mathcal{D}^i_B = e\int d^3x\, x^i \langle B,s|J^0_{\rm em}|B,s\rangle = \mathcal{D}_B \langle s| \sigma^i|s \rangle,
\eeq
where $|B,s\rangle$ is a baryonic state and the last equality defines
$\mathcal{D}_B$ requiring the EDM vector to be proportional to the
spin (since it is the only physical vector intrinsic to the
baryon). 

We call the subleading correction we are about to compute $\DeltaN^i$ following \eqref{notationEDM}, since it will correspond to the isoscalar contribution.
Of course it will still obey the relation
\beq
\DeltaN^i = \DeltaN \langle s |\sigma^i|s\rangle.
\eeq
Our aim now is to compute the value of $\DeltaN$. As we already
mentioned, only the Abelian part of the current \eqref{Jem} will give
contributions to it, so for our purpose, the perturbed current
effectively reads 
\beq
\delta J^0_{\rm em} =-\kappa\left[k(z)\delta\hF^{0z}\right]^{+\infty}_{-\infty}= \frac{\kappa}{\Nc}\left[k(z)\de_z\delta\hA^0\right]^{+\infty}_{-\infty}=\Upsilon \frac{\kappa}{\Nc} \left[k(z)\de_z\mathcal{M}\right]^{+\infty}_{-\infty} \xc.
\eeq
Plugging this expression into the EDM formula yields:
\beq
\DeltaN^i = e\Upsilon\frac{\kappa}{\Nc}\int d^3x\, x^i  \langle {\rm N}
|\left[k(z)\de_z\mathcal{M}\right]^{+\infty}_{-\infty}\xc|{\rm N}\rangle.
\eeq
Since we approximate the massive moduli by their classical values, we
can just keep the angular velocity in the expectation value. We
further switch to spherical coordinates and integrate over $d\Omega$: 
\beq\label{deltaNi}
\DeltaN^i = e\Upsilon\frac{\kappa}{\Nc}\frac{4\pi}{3}\int dr\, r^3
\left[k(z)\de_z\mathcal{M}\right]^{+\infty}_{-\infty} \langle {\rm N}
|\chi^i|{\rm N}\rangle.
\eeq
Now, making use of (\ref{Jchi}) (setting again $\rho = \rho_{\rm cl}$)
and writing $J^k \equiv \frac{1}{2}\sigma^k$ we finally obtain: 
\beq
\DeltaN^i = \frac{ecm\theta}{192\pi^3\kappa^2}\sqrt{\frac{5}{6}}\int
dr\, r^3 \left[k(z)\de_z\mathcal{M}\right]^{+\infty}_{-\infty} \langle
{\rm N} |\sigma^i|{\rm N}\rangle.
\eeq
To make a prediction for $\DeltaN$ we use, other than $\Nc=3$, the
most common parameter choices for the Sakai-Sugimoto model, i.e.: 
\beq
\kappa = 0.00745\quad ;\quad M_{\KK} = 949\;\text{MeV} \quad; \quad m = 2.92\;\text{MeV}
\eeq
The quark mass $m$ is chosen such that it correctly reproduces
$m_{\pi}= 135\;$MeV in the GMOR relation $4cm = f_{\pi}^2 m_{\pi}^2$,
and it turns out being a physically reasonable value that lies in
between those of the up and down quark masses.
The pion decay constant is given in Ref.~\cite{Sakai:2004cn} in terms of
$\kappa$: 
\beq
f_{\pi}^2 = 4\frac{\kappa}{\pi}.
\eeq
With these choices, and restoring factors of $M_{\KK}$ by simple
dimensional analysis, our prediction is 
\beq
\DeltaN = -4.6\times 10^{-17}\theta \text{ $e\cdot$cm}
\eeq

It is possible to repeat the computation for different values of
$\lambda$ in order to extract the scaling of the EDM in the large
$\lambda$ limit. Of course there is a limitation to how large we
can take $\lambda$, since for $\lambda \rightarrow\infty$ the
instanton becomes pointlike and the precision of the numerical
solution is lost.
Nevertheless, we manage to reach $\lambda = 10^{3.5}$ while keeping a
trustable solution.  

\begin{figure}
	\centering
	\includegraphics[width=0.65\linewidth]{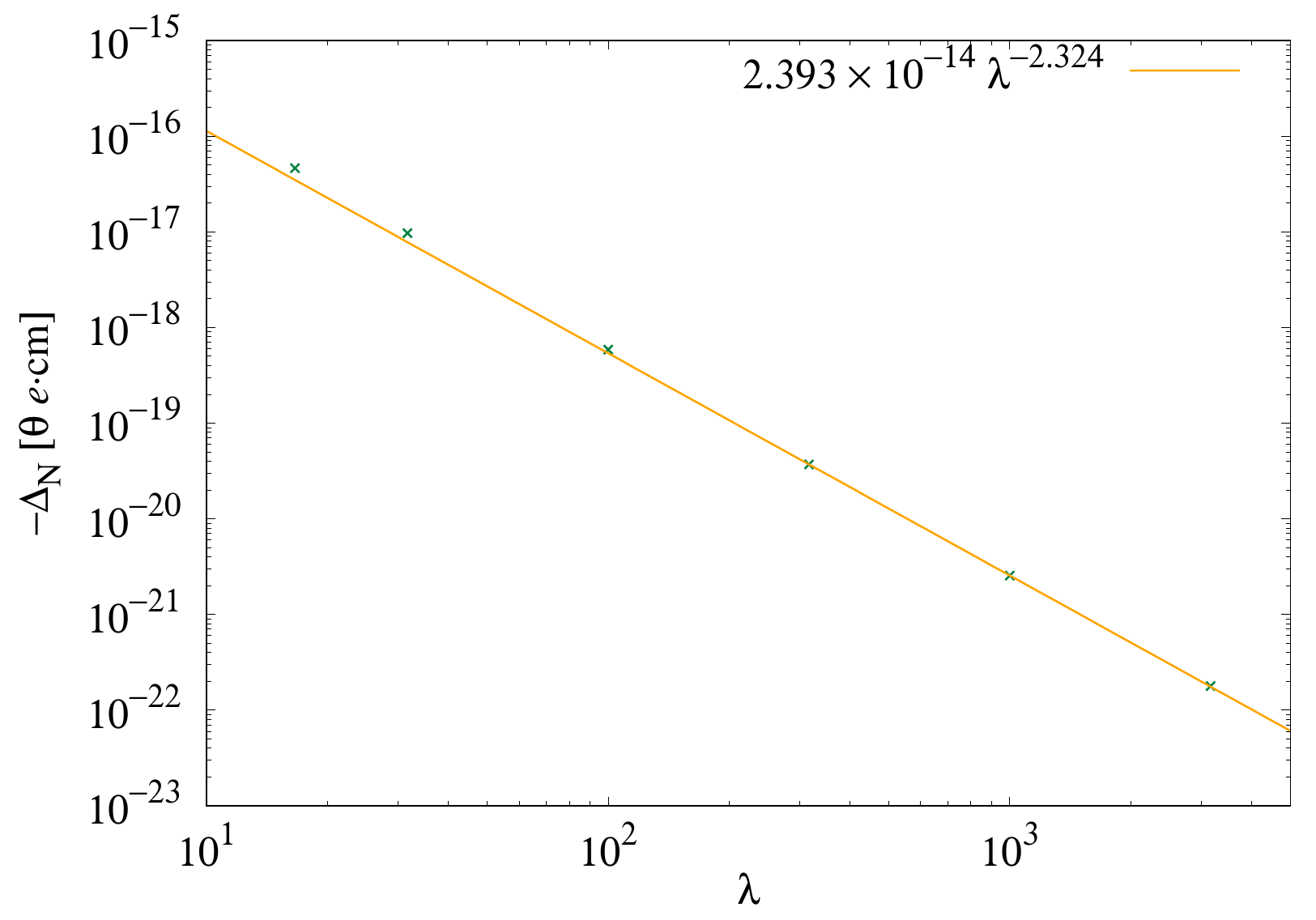}
	\caption{\small The logarithmic plot of $\DeltaN$ for increasing values of $\lambda$, starting with the phenomenological one. As can be seen, the $\lambda$ dependency tends to a definite power law in the large $\lambda$ limit.}
          \label{scaling} 
\end{figure}

The result we obtain for the scaling at large $\lambda$ is 
\beq
\DeltaN \sim -2.393 \times 10^{-14} \lambda^{-2.324} \theta \text{ $e\cdot$cm} ,
\eeq
see Fig.~\ref{scaling}.
Note that this contribution is consistently suppressed with respect to the isovectorial one, that scales as $\lambda^{-2}$ \cite{Bartolini:2016dbk}, but not strongly, which allows to obtain the correct order of magnitude with extrapolation to phenomenological $\lambda$. 
\section{From the nucleons to the deuteron}
\label{cinque}

Computing the EDM of the deuteron requires us to have $B=2$ quantum
states: Of course we need in particular the ground state of that
topological sector. There are at least two different consistent ways
of obtaining such state, following from the non-commutativity of the
two large-$\Nc$ and large-$\lambda$ limits.  
Nonetheless, at leading order, our computation is not dependent on
such details, so it yields the same result no matter how we build the
Sakai-Sugimoto deuteron state as long as it has the correct quantum
numbers. 

A few considerations on such numbers: we know from phenomenology that
the ground state is in the isospin singlet, spin triplet configuration
$(I=0,J=1)$, and its orbital wave function is mostly composed of
the $L=0$ state, with a small part of the $L=2$ one. We will assume $L=0$ from
now on, since from the holographic point of view the $L=2$ component
has to be suppressed by powers of $\lambda^{-1}$: this can simply be
understood by considering that the $L=2$ component is geometrically
realized by the two nucleons spinning around an axis orthogonal to
their separation. In this configuration, we can estimate the moment of
inertia for rotations around this axis as $2M_B R^2$: The separation
between the nucleons' cores is of order $R\sim \mathcal{O}(1)$, as
verified in Ref.~\cite{Baldino:2017mqq}, while the leading order of the
baryon mass is given by $M_B = 8\pi^2 \kappa$. Hence it is of order
$\lambda$ and so is the moment of inertia. 

On the other hand, the $L=0$ configuration involves no other angular
momentum than the spin of the nucleons, as it can be thought of as
the two spins lying along the separation between the nucleons and
pointing in the same direction. Thus the moment of inertia for this
angular momentum is given by the sum of the ones of the single
solitons, each amounting to $4\pi^2 \kappa \rho^2$. Since the
classical value of the size is given by Eq.~\eqref{classicalrhoz}, this
moment of inertia does not scale with $\lambda$, hence the $L=0$
component dominates the orbital wave function once the large-$\lambda$
limit is taken. 
\begin{figure}
	\centering
	\includegraphics[width=12cm]{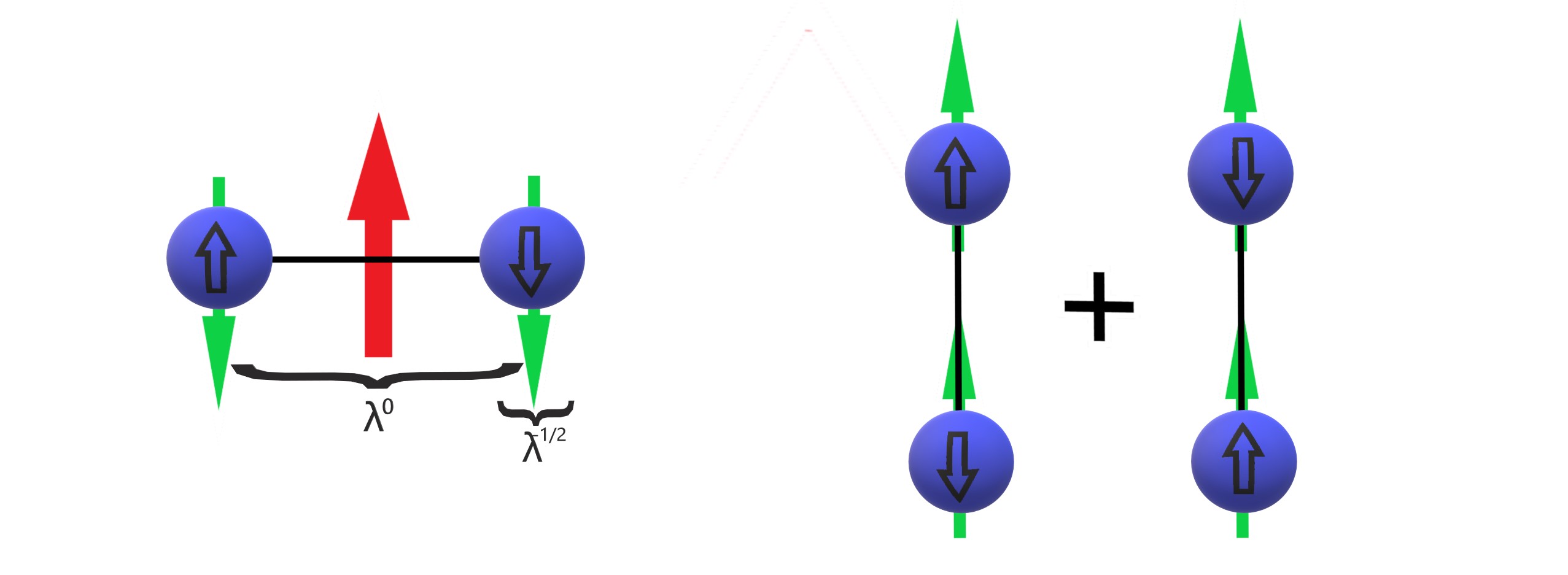}
	\caption{\small Configuration of the two solitons in the $L=2$
          (left) and $L=0$ (right) sectors: The arrows denote the
          directions of spatial angular momentum (red), single soliton
          spin (green) and single soliton iso-orientation (on the
          soliton). The size of each soliton is of order
          $\lambda^{-1/2}$, while the separation between them is of
          order $\lambda^{0}$. In the quantum ground state, each
          soliton is in a superposition of opposite isospin
          direction.}\label{solmass}. 
\end{figure}

\subsection{Deuteron EDM}

The deuteron is shaped by placing two solitons at the distance $R$
that minimizes the nucleon-nucleon potential, and assigning to each of
them the $SU(2)$ orientation described respectively by the matrices $B,C$: 
\beq
\mA = B\mA^{\rm cl}_{(1)}\left(\vec{x}+\frac{\vec{R}}{2},z\right)B^{\dagger} + C\mA^{\rm cl}_{(2)}\left(\vec{x}-\frac{\vec{R}}{2},z\right)C^{\dagger}.
\eeq
The two approaches in the construction of the deuteron treat
the moduli of $SU(2)$ differently: The solitons are either treated as
spinning independently, or as having a locked relative
orientation. Since we are interested in the Abelian part of the
current, such details will not play any role, as the $SU(2)$ moduli
will only enter the computation via the total angular momentum. 
The full EDM can be computed as two separate contributions:
\beq
\calDD^i =  e\int d^3x\left( x^i-x_0^i\right) \langle {\rm
  D},s|\tr\left(\delta J^0_{V}\tau^3\right)+ \frac{1}{\Nc}\delta
\widehat{J}^0_V|{\rm D},s\rangle = \left(\dD + \DeltaD \right)\langle j|\sigma^i|j\rangle.
\eeq
In the following sections we will show that, in both approaches to the
deuteron, we obtain the simple results 
\beq
&&\dD =0, \\
&&\DeltaD =2\DeltaN.
\eeq
The $SU(2)$ part of the electromagnetic charge density comes in the form
\beq\label{JSU2}
\tr\left(\delta J^0_V\tau^3\right) = K\kappa \left[k(z)\de_zW^{(1)}\hat{r}_1\cdot\tr\left(B\vec{\tau}B^{\dagger}\tau^3\right)+k(z)\de_zW^{(2)}\hat{r}_2\cdot\left(\ C \vec{\tau}C^{\dagger}\tau^3\right)\right]^{+\infty}_{-\infty}.
\eeq
The complete field strength $\delta F^{0z}$ would also include a term
of the form
$[\delta A^0_{(1)}$+$\delta A^0_{(2)}$, $A^z_{(1)}$ + $A^{z}_{(2)}]$
but it can easily be checked to vanish,  
since $\delta A^0$ and $A^z$ share the same group structure
$f(r,z)\vec{x}\cdot \A\vec{\tau}\A^{-1}$. 

The new $U(1)$ part reads
\beq\label{JU1}
\delta \widehat{J}^0_{\rm em}\equiv\frac{1}{\Nc}\delta \widehat{J}^{0}_{V} = \Upsilon\frac{\kappa}{\Nc}\left[k(z)\de_z\mathcal{M}^{(1)}(\hat{r}_1\cdot\vec{\chi}^{(1)}) + k(z)\de_z\mathcal{M}^{(2)}(\hat{r}_2\cdot\vec{\chi}^{(2)})\right]^{+\infty}_{-\infty}.
\eeq
In both equations, we have defined
$\hat{r}_1 =  \frac{\vec{x}+\frac{\vec{R}}{2}}{|\vec{x}+\frac{\vec{R}}{2}|}$ and $\hat{r}_2 = \frac{\vec{x}-\frac{\vec{R}}{2}}{|\vec{x}-\frac{\vec{R}}{2}|} $.

\subsection{Approach 1}

In this approach, given in Ref.~\cite{Baldino:2017mqq}, the deuteron state
$|{\rm D}\rangle$ is obtained by quantizing the $B=2$ zero modes
manifold: the massless $SU(2)\times SU(2)$ moduli corresponding to
global iso- and spatial rotations are given by the matrices
$U\equiv u_4+iu_k\tau^k$ and $E\equiv e_4 +ie_k\tau^k$. They can be
related to the single soliton moduli $B$ and $C$ via the embedding
law:  
\beq
&&B=UE^{\dagger},\\
&&C=iU\tau^3 E^{\dagger},\label{Cembed}
\eeq
where the factor $i\tau^3$ in Eq.~\eqref{Cembed} is present because the
relative orientation of the nucleons is not a massless modulus: the
nucleon-nucleon potential is found to be a function of the moduli
$(\rho_1,\rho_2,Z_1,Z_2, B^{\dagger}C)$, hence the
$i\tau^3$ factor selects the attractive channel, performing a relative
rotation in isospin space of $\pi$ around an axis orthogonal to the
separation between nucleons. 

The found deuteron state can be written in terms of the global moduli
$e_I$: 
\beq
\langle e_I,u_I|{\rm D}\rangle = \frac{1}{\pi^2}\left(2\left(e_3^2 + e_4^2\right) -1 \right),
\eeq
but it is more useful to write it using the single soliton moduli
$b_I,c_I$:  
\beq\label{Dbc}
\langle b_I,c_I|{\rm D}\rangle  =\frac{1}{\pi^2} \left(b_4c_3-b_3c_4+b_1c_2 -b_2 c_1\right).
\eeq
As a first step, we show that the dipole moment of Eq.~\eqref{JSU2}
vanishes on the deuteron state. We have to compute the quantity 
\beq
\dD^i = eK\kappa\langle {\rm D}|\int d^3x\, x^i
\left[k(z)\left(\de_zW^{(1)}\hat{r}_1\cdot\tr\left(B\vec{\tau}B^{\dagger}\tau^3\right)+\de_z{W}^{(2)}\hat{r}_2\cdot\left(C
  \vec{\tau}C^{\dagger}\tau^3\right)\right)\right]^{+\infty}_{-\infty}|{\rm
  D}\rangle.\nonumber
\eeq
To begin with, we note that the two integrals of $\dot{W}^{(1)}$ and
$\dot{W}^{(2)}$ give the same result, since it is sufficient to
perform separately the change of variables $\vec{x}\rightarrow \vec{x}
\mp\frac{\vec{r}}{2}$ to make them explicitly the same integral. Then
we note that Eq.~\eqref{Dbc} is antisymmetric under the exchange of $b_I$
with $c_I$. Thus to obtain the full result, we only need to compute 
\beq
\langle{\rm D} |\tr\left(B\tau^iB^{\dagger}\tau^3\right)|{\rm D}\rangle,
\eeq
which turns out to vanish for every $i=1,2,3$. Hence we conclude that the
$SU(2)$ part of the current does not contribute to the deuteron EDM:
this is in line with what we expected, a result proportional to the
total isospin, which is zero for the deuteron. In principle one could
expect the contributions of the two nucleons to cancel each other, as
the classical picture of a neutron with $I_3=-\frac{1}{2}$ and a
proton with $I_3=+\frac{1}{2}$ would suggest: the fact that each
contribution vanishes on its own instead is due to the fact that the
quantum state Eq.~\eqref{Dbc} does not assign a definite $I_3$ to each
nucleon, but both are in an equally probable superposition of neutron
and proton states (as shown in fig.~\ref{solmass}), hence the average
$I_3$ of each soliton vanishes. 

Now we turn to the computation of $\DeltaD^i$:
\beq
\DeltaD^i= e\Upsilon\frac{\kappa}{\Nc}\langle {\rm D} |\int d^3x\,
x^i\left[k(z)\de_z\mathcal{M}^{(1)}(\hat{r}_1\cdot\vec{\chi}^{(1)}) +
  k(z)\de_z\mathcal{M}^{(2)}(\hat{r}_2\cdot\vec{\chi}^{(2)})\right]^{+\infty}_{-\infty}|{\rm
  D}\rangle.
\eeq
As before, the integrals can be evaluated separately, and each of them
reproduce the result of Eq.~\eqref{deltaNi}, so we are left with 
\beq
\DeltaD^i=e\Upsilon\frac{\kappa}{\Nc}\frac{4\pi}{3}\int dr\, r^3
\left[k(z)\dot{\mathcal{M}}\right]^{+\infty}_{-\infty} \langle {\rm D}
|\chi^i_{(1)}+\chi^i_{(2)}|{\rm D}\rangle.
\eeq
By making use of (\ref{Jchi}) we trade the angular velocities for the angular momenta:
\beq
\DeltaD^i=\frac{e\Upsilon}{3\pi\rho^2\Nc}\int dr\, r^3
\left[k(z)\de_z\mathcal{M}\right]^{+\infty}_{-\infty} \langle {\rm D}
|J^i_{(1)}+J^i_{(2)}|{\rm D}\rangle.
\eeq
The last step is to use the fact that $L=0$, so effectively
$J_{\rm D}^i = J_{(1)}^i+J_{(2)}^i$, and thus we obtain the
aforementioned result  
\beq
\DeltaD^i=\frac{ecm\theta}{96\pi^3\kappa^2}\sqrt{\frac{5}{6}}\int dr\, 
r^3 \left[k(z)\de_z\mathcal{M}\right]^{+\infty}_{-\infty} \langle {\rm
  D} |J_{\rm D}^i|{\rm D}\rangle = 2\DeltaN \langle {\rm D} |J_{\rm
  D}^i|{\rm D}\rangle.\label{eq:DeltaDi}
\eeq
which is the full result for the EDM of the deuteron:
\beq\label{Ddeuton}
\calDD = 2\DeltaN = -0.92 \times 10^{-16}\theta\text{ $e\cdot$cm}
\eeq

\subsection{Approach 2}

Another possible setup is the one adopted in Ref.~\cite{Kim:2009sr}. Since
the two solitons are placed at a distance much greater than their
size, they can be treated as independent identical particles. 
Since each of them is quantized as a fermion, we can build the global
wave function $|{\rm D}\rangle$ as an antisymmetric combination of the two
single soliton states with $SU(2)$ quantum numbers $l=1$,
$|{\rm N}\rangle =|l/2=\frac{1}{2}, m_s, m_i\rangle$. Antisymmetry in
the $I_3$ quantum number leads us to 
\begin{equation}
|{\rm D},m_j\rangle =
\frac{1}{\sqrt{2}}\left(|{\rm p},m_s\rangle|{\rm n},m_s\rangle-|{\rm
  n},m_s\rangle|{\rm p},m_s\rangle\right), 
\end{equation}
with $m_j = 2m_s$.
This configuration still does not assign a definite third component of
the isospin to any of the two solitons (it is still of the type
illustrated on the right-hand side of Fig.~\ref{solmass}), so the argument
for the vanishing of $\dD$ we used in the previous section is still
valid here. 

It is also still true that $J_{\rm D}^i=J_{(1)}^i + J_{(2)}^i$ so
Eq.~\eqref{Ddeuton} also holds its validity.

\section{Conclusion} 
\label{sei}

Using the holographic model of Witten-Sakai-Sugimoto, we were able to
extend the computation of the EDMs of baryons to the isoscalar
part. It turns out to be of a comparable magnitude with the
isovectorial one, once extrapolation to phenomenological values of the
parameters of the model is performed, despite it being a subleading
correction in $\lambda^{-1}$ and $\Nc^{-1}$. In particular, we observe
the scalings $\DeltaN/d_{N}\sim \mathcal{O}(\lambda^{-1}\Nc^{-2})$. 

Using the deuteron description emerging from the same model and the
results for the EDMs of nucleons, we were able to estimate the EDM of
the deuteron bound state, obtaining a value close to the estimate
given in Ref.~\cite{Lebedev:2004va}: even if this numerical closeness
may be regarded as an accident, considering the many approximations 
implicit in our computations (and the lack of the inclusion of
two-body contributions which are expected to give comparable EDMs),
it is still remarkable that we obtain the correct order of magnitude
and sign, despite this term being formally subleading in $\lambda$ and
$\Nc$ before phenomenological extrapolation of the parameters.
This can be ultimately traced back to the known fact that the perturbative regime in the Sakai-Sugimoto model is not well established at phenomenological values of the model parameters, especially for the baryonic sector. In this sense, it is clear that our result for the deuteron EDM can receive significant corrections at subsequent orders in this perturbative expansion (as happens explicitly for the single nucleons) and is thus to be regarded as order of magnitude estimates of the EDMs. While it is not really significant in this sense to change the estimates of the single nucleon EDMs previously obtained in Refs.~\cite{Bartolini:2016dbk} and \cite{Bartolini:2016jxq}, as the exact value can be further modified with higher order corrections, it is indeed relevant for the newly computed deuteron EDM, being the leading order and thus establishing the order of magnitude and sign for the quantity within this model. Moreover, we stress that unlike the single nucleon EDMs, the deuteron EDM receive corrections only from terms in the perturbative expansion that can contribute to the isoscalar charge density $\delta \widehat{J}^{0z}$: extending the mechanism that generates source terms for the perturbations from the mass term in the equations of motion, it is clear that the next contribution to the isoscalar current would arise at NNNLO, as the one at NNLO is isovectorial.

Two-body terms can be divided into two conceptually different classes:
polarization terms ($\calDD^{\rm (pol)}$) and exchange terms
($\calDD^{\rm (exc)}$). The first ones account for P-wave components
in the wave function of the deuteron, and pion-nucleon coupling
$\bar{g}_{\pi NN}^{(1)}$. The second class arises from the exchange of
currents between the nucleons, and can potentially receive
contributions from both the isospin-preserving, CP-breaking
pion-nucleon couplings $\bar{g}_{\pi NN}^{(0)}$ and
$\bar{g}_{\pi{\rm NN}}^{(1)}$. 
The term that dominates, however, is expected to be the polarization
one, and in the exchange term the bigger role is played by pieces
proportional to $\bar{g}_{\pi NN}^{(1)}$. 
However, in the setup we employed, we only expect two-body
contributions to arise from $\bar{g}_{\pi NN}^{(0)}$, since we did not
include isospin-breaking terms in the quark mass matrix, we lose
all the larger pieces of this two-body term. 

To be fully self consistent we only need to account for the exchange
term that picks up $\bar{g}_{\pi NN}^{(0)}$: Conceptually, one would
need to perturb the full two-soliton configuration, and look for
$\theta$-induced perturbations of the soliton tail. This looks like
an overly-hard task, but it is reasonable to expect that such term is
subleading in $\lambda^{-1}$, being the outcome of the perturbation of
a solitonic tail (which can be regarded as a perturbation to the
soliton core) induced by a perturbation of the cores (that is, the
$\theta$-induced perturbations we found).

\section*{Acknowledgments}
 We thank A.~Cotrone and I.~Basile for useful discussions and especially F.~Bigazzi and P.~Niro
 for various discussions and collaboration at the initial stage of
 this work. 
 This work is supported by the INFN special project grant ``GAST
 (Gauge and String Theory)''.
 The work of S.~B.~G.~is supported by the National Natural Science
 Foundation of China (Grant No.~11675223). 
 S.~B.~G.~thanks the Outstanding Talent Program of Henan University
 for partial support. 

\appendix

\section{Explicit equations of motion}\label{A}

Here we provide the equations of motion to be solved for every group
structure of the Ansatz we employed. 
The function $W(r,z)$ is defined by Eq.~\eqref{defW}.
The functions $\beta, \gamma, \delta$ appear with only the first
derivative with respect to $z$, while the functions $D,F,G$ appear
with only the first derivative with respect to $r$. All the other
functions appear with all the derivatives up to second order with
respect to both coordinates. 
Note that every function has a definite parity under $z\rightarrow-z$,
so the boundary condition at infinity for the $z$ coordinate can
be imposed either at $z=+\infty$ or $z=-\infty$: The equations will
take care of the behavior of the functions on the other side of the
$z$ axis.
The boundary conditions at $z=0$ can instead be guessed from the
parity of each function: $\beta,\gamma,E,G,H,I$ are even, while
$\delta, B,C,D,F$ are odd. 
The boundary conditions we impose are thus:
\bea\label{BC}
&&\beta'(0,z)=\gamma'(0,z)=\delta'(0,z)=0, \nonumber \\
&&\beta(+\infty,z) = \gamma (+\infty,z) = \delta(+\infty,z) = 0, \nonumber \\
&&\beta(r,\infty) = \gamma (r,\infty) = \delta(r, \infty) = 0, \nonumber \\
&&B'(0,z) = C'(0,z) =E'(0,z)=H'(0,z)=I'(0,z)=0,  \nonumber \\
&&X(+\infty,z)=0,\quad\qquad \qquad \qquad{\rm for}\qquad X=B,\dots,I, \nonumber \\
&&B(r,0)=C(r,0)=D(r,0)=F(r,0)=0, \nonumber \\
&&\dot{E}(r,0)=\dot{G}(r,0)=\dot{H}(r,0)=\dot{I}(r,0)=0, \nonumber \\
&&X(r,\infty)=0,\qquad\qquad \qquad\qquad{\rm for}\qquad X=B,\dots,I.
	\eea
\begin{itemize}
 \item{$\xc\xt$}
{\footnotesize
	\beq
\nonumber
	&&\dot{B'}-\frac{1}{r}\dot{B}+\dot{C'}-\frac{1}{r}\dot{C}+\dot{F'}+\frac{2}{r}\dot{F}-\beta''-\frac{2}{r}\beta'+\frac{6}{r^2}\beta\\\nonumber
	&&+\frac{2}{\xr}\left[r\dot{B}+2r\dot{C}+ rE'-rG'+z\dot{G}-3G-z\dot{H}+2z\dot{I}-3I-6\beta-2z\delta'+2\frac{z}{r}\delta \right]\\\nonumber
	&&+\frac{4}{(\xr)^2}\left[-2zrC -zrD+r^2E+(r^2-\rr)G+\rr H+2(r^2-\rr)I+2\xx\beta+r^2\gamma + zr\delta \right]\\\nonumber
	&&+\frac{2r^2}{\xr}\left[z\dot{W}-\frac{\rr}{\xr}W\right]\\
	&&-\frac{\pi}{16} \frac{\rr}{(\rr+r^2)^{3/2}}\frac{r}{k(z)}\sin\left(\frac{\pi}{\sqrt{1+\frac{\rr}{r^2}}}\right)= 0,
\eeq
}
	\item{$\ct $}
{\footnotesize 
	\beq\nonumber
	&&\frac{1}{r}\dot{B}+\frac{1}{r}\dot{C}+\dot{D'}+\frac{2}{r}\dot{D}-\frac{2}{r^2}\beta-\gamma''-\frac{2}{r}\gamma'\\\nonumber
	&&+\frac{2}{\xr}\left[-r\dot{B}+2z\dot{E}-rE'-2E-z\dot{G}+rG'+3G+z\dot{H}+I+2\beta+2z\delta'+2\frac{z}{r}\delta  \right]\\\nonumber
	&&+\frac{4}{(\xr)^2}\left[ (2z^2+r^2)\gamma-zr\delta+ zrD+(r^2-2\rr )E+(\rr-r^2 )G-\rr H\right]\\
	&&+\frac{2r^2}{\xr}\left[-z\dot{W}+\frac{\rr}{\xr}W\right] = 0,
	\eeq
}
 \item{$\epsilon^{abc}\chi^a \hat{r}^b\tau^c $}
{\footnotesize 
	\beq\nonumber
	&&-\dot{E'}+\dot{G'}+\frac{2}{r}\dot{G}+\frac{1}{r}\dot{H}+\frac{1}{r}\dot{I}-\delta''-\frac{2}{r}\delta'+\frac{2}{r^2}\delta\\\nonumber
	&&+\frac{2}{\xr}\left[ -z\dot{B}-C+z\dot{D}-rD'-3D-r\dot{E}-r\dot{H}+2\frac{z}{r}\beta -2z\gamma'-2\delta \right]\\\nonumber
	&&+\frac{4}{(\xr)^2}\left[ \rr B+(r^2-\rr) D -zrE +zrG +zr\gamma +(2z^2+r^2)\delta \right]\\
	&&+\frac{r}{\xr}\left[-(z^2-r^2)\dot{W}+\frac{2\rr}{\xr}zW\right] = 0,
	\eeq
}
 \item{$\chi^i \xt $}
{\footnotesize 
	\beq \nonumber
	&&h(z)\left[-B''-\frac{1}{r}B'+\frac{1}{r^2}B+\frac{1}{r}C'-\frac{1}{r^2}C+\frac{1}{r}F'\right]+k(z)\left[\frac{1}{r}\dot{\beta}-\ddot{B}\right]+2z\left[\frac{1}{r}\beta-\dot{B}\right] \\ \nonumber
	&&+\frac{2h(z)}{\xr}\left[-2B+C+rD'+2D-zE'+zG'+\frac{z}{r}G   +\frac{z}{r}I\right]\\ \nonumber
	&&+\frac{4h(z)}{(\xr)^2}\left[ \xx B+(\rr-r^2)D +zrE-zrG \right]\\ \nonumber
	&&+\frac{2k(z)}{\xr}\left[z\dot{\delta}+\delta+r\dot{\gamma}   \right] + \frac{4z}{\xr}\left[ r\gamma +z\delta \right]\\ \nonumber
	&&+\frac{4k(z)}{(\xr)^2}\left[ -z^2\delta +\rr \delta -zr\gamma \right]\\
	&&+h(z)\frac{W}{(\xr)^2}2zr\rr=0,
	\eeq
}
 \item{$\xc \tau^i $}
{\footnotesize 
	\beq \nonumber
	&&h(z)\left[+\frac{1}{r}B'-\frac{1}{r^2}B-C''-\frac{1}{r}C'+\frac{1}{r^2}C+\frac{1}{r}F'\right]+k(z)\left[\frac{1}{r}\dot{\beta}-\ddot{C}\right]+2z\left[\frac{1}{r}\beta-\dot{C}\right] \\ \nonumber
	&&+\frac{2h(z)}{\xr}\left[ -rB'-B-rC'-2C-rD'-2D+2zE'-rF'-2F+\frac{z}{r}G-3\frac{z}{r}H+2zI'+4\frac{z}{r}I  \right]\\ \nonumber
	&&+\frac{4h(z)}{(\xr)^2}\left[ (\xx-\rr)B+(4z^2+r^2-2\rr)C+(\xx-\rr)D -3zrE+(\xx-\rr) F-3zrI\right]\\ \nonumber
	&&+\frac{2k(z)}{\xr}\left[ -2r\dot{E} -2r\dot{I}-r\dot{\beta}-r\dot{\gamma}-\delta  \right] + \frac{4z}{\xr}\left[-rE-rI-r\beta-r\gamma  \right]\\ \nonumber
	&&+\frac{4k(z)}{(\xr)^2}\left[ r^2C+zrE+zrI \right]\\
	&&+h(z)\frac{W}{(\xr)^2}2zr(\xx-\rr)=0,
	\eeq
}
 \item{$\hat{r}^i \ct $}
{\footnotesize 
	\beq \nonumber
	&&h(z)\left[\frac{1}{r}B'-\frac{1}{r^2}B+\frac{1}{r}C'-\frac{1}{r^2}C-\frac{2}{r^2}F\right]+k(z)\left[\dot{\gamma'}-\ddot{D}\right]+2z\left[\gamma'-\dot{D}\right] \\ \nonumber
	&&+\frac{2h(z)}{\xr}\left[-rB'+B+2C+zE'+2F+zH'+\frac{z}{r}H-\frac{z}{r}I   \right]\\ \nonumber
	&&+\frac{4h(z)}{(\xr)^2}\left[\rr B+ \xx D \right]\\ \nonumber
	&&+\frac{2k(z)}{\xr}\left[ 2r\dot{E}-2r\dot{G}-z\dot{\delta}-\delta+r\delta'  \right] + \frac{4z}{\xr}\left[ rE-rG-z\delta \right]\\ \nonumber
	&&+\frac{4k(z)}{(\xr)^2}\left[r^2D-zrE+zrG+zr\gamma+(z^2-\rr)\delta  \right]\\
	&&+h(z)\left[-\frac{W'}{\xr}2zr^2-\frac{ W}{(\xr)^2}2zr\xx\right]=0,
	\eeq
}
 \item{$\epsilon^{iab}\chi^a\tau^b $}
{\footnotesize 
	\beq \nonumber
	&&h(z)\left[-E'' -\frac{1}{r}E'-\frac{1}{r}G'-\frac{3}{r^2}H-\frac{3}{r^2}I\right]+k(z)\left[-\frac{1}{r}\dot{\delta}-\ddot{E}\right]+2z\left[-\frac{1}{r}\delta-\dot{E}\right] \\ \nonumber
	&&+\frac{2h(z)}{\xr}\left[ -3\frac{z}{r}C-zD'-\frac{z}{r}D-E+4H  \right]\\ \nonumber
	&&+\frac{4h(z)}{(\xr)^2}\left[ -zrB+zrD+(\xx-\rr)E-r^2H \right]\\ \nonumber
	&&+\frac{2k(z)}{\xr}\left[ -z\dot{\gamma}-\gamma  \right] + \frac{4z}{\xr}\left[ -z\gamma \right]\\ \nonumber
	&&+\frac{4k(z)}{(\xr)^2}\left[ (z^2-\rr)\gamma \right]\\
	&&+h(z)\frac{W}{\xr}(z^2-r^2)=0,
	\eeq
}
\item{$\hat{r}^i\xc\xt $}
{\footnotesize 
	\beq \nonumber
	&&h(z)\left[B''-\frac{3}{r}B'+\frac{3}{r^2}B+C''-\frac{3}{r}C'+\frac{3}{r^2}C-\frac{2}{r}F'+\frac{6}{r^2}F\right]\\\nonumber
	&&\qquad \qquad+k(z)\left[\dot{\beta'}-\frac{2}{r}\dot{\beta}-\ddot{F}\right]+2z\left[\beta'-\frac{2}{r}\beta -\dot{F}\right]\\ \nonumber
	&&+\frac{2h(z)}{\xr}\left[2rB'-2B+3rC'-3C+rF'-4F-zG'+2\frac{z}{r}G-zH'+2\frac{z}{r}H   \right]\\ \nonumber
	&&+\frac{4h(z)}{(\xr)^2}\left[ r^2C+r^2D+(\xx+\rr) F+zrG+zrI \right]\\ \nonumber
	&&+\frac{2k(z)}{\xr}\left[  2r\dot{G}+2r\dot{I}+r\dot{\beta}-r\delta'+\delta \right] + \frac{4z}{\xr}\left[rG+rI+r\beta  \right]\\ \nonumber
	&&+\frac{4k(z)}{(\xr)^2}\left[-r^2C-r^2D-zrG-zrI  \right]\\
	&&+h(z)\frac{W'}{\xr}2zr^2=0,
	\eeq
}
\item{$\hat{r}^i\epsilon^{abc}\chi^a\hat{r}^b\tau^c $}
{\footnotesize 
	\beq \nonumber
	&&h(z)\left[-E''+\frac{ 1 }{ r }E'-\frac{1}{r}G'+\frac{2}{r^2}G+\frac{1}{r}H'-\frac{2}{r^2}H+\frac{1}{r}I'-\frac{2}{r^2}I\right]\\\nonumber 
	&&\qquad \qquad +k(z)\left[\dot{\delta'}-\frac{1}{r}\dot{\delta}-\ddot{G}\right]+2z\left[\delta'-\frac{1}{r}\delta-\dot{G}\right] \\ \nonumber
	&&+\frac{2h(z)}{\xr}\left[ -zB'+\frac{z}{r}B-zD'+\frac{z}{r}D-rE'+2\frac{z}{r}F-2G-rH'+3H  \right]\\ \nonumber
	&&+\frac{4h(z)}{(\xr)^2}\left[ -zrB+zrD+\xx G+(\rr - r^2)H \right]\\ \nonumber
	&&+\frac{2k(z)}{\xr}\left[ 2r\dot{D}-r\gamma'  \right] + \frac{4z}{\xr}\left[  rD\right]\\ \nonumber
	&&+\frac{4k(z)}{(\xr)^2}\left[-zrD-r^2E+r^2G+zr\delta  \right]\\
	&&+h(z)\left[-\frac{W'}{\xr}r(z^2-r^2) -\frac{ W}{(\xr)^2}2r^2\rr\right]=0,
	\eeq
}
	\item{$\epsilon^{iab}\chi^a \hat{r}^b\xt $}
{\footnotesize 
	\beq \nonumber
	&&h(z)\left[-H''-\frac{2}{r}H'+\frac{6}{r^2}H\right]+k(z)\left[-\ddot{H}\right]+2z\left[-\dot{H}\right] \\ \nonumber
	&&+\frac{2h(z)}{\xr}\left[-rE'+rG'+2G-6H+3I   \right]\\ \nonumber
	&&+\frac{4h(z)}{(\xr)^2}\left[ zrB+r^2E+(\rr-r^2)G+(z^2+2r^2)H \right]\\ \nonumber
	&&+\frac{2k(z)}{\xr}\left[ r\dot{\delta}  \right] + \frac{4z}{\xr}\left[  r\delta\right]\\ \nonumber
	&&+\frac{4k(z)}{(\xr)^2}\left[ -zr\delta \right]\\
	&&+h(z)\frac{W}{(\xr)^2}2r^2\xx =0,
	\eeq
}
	\item{$\xc \epsilon^{iab}\hat{r}^a\tau^b $}
{\footnotesize 
	\beq \nonumber
	&&h(z)\left[-I''-\frac{2}{r}I'+\frac{6}{r^2}I\right]+k(z)\left[-\ddot{I}\right]+2z\left[-\dot{I}\right] \\ \nonumber
	&&+\frac{2h(z)}{\xr}\left[ -zB'+\frac{z}{r}B-3zC'+3\frac{z}{r}C-zF'-H-4I  \right]\\ \nonumber
	&&+\frac{4h(z)}{(\xr)^2}\left[ zrB-zrC-zrD+r^2E+r^2H+(z^2+2r^2-\rr)I \right]\\ \nonumber
	&&+\frac{2k(z)}{\xr}\left[2r\dot{C}-z\dot{\beta}-2\beta   \right] + \frac{4z}{\xr}\left[rC-z\beta  \right]\\ \nonumber
	&&+\frac{4k(z)}{(\xr)^2}\left[  -zrC+r^2E+r^2I+(\xx-\rr)\beta +r^2\gamma\right]\\
	&&+h(z)\frac{W}{(\xr)^2}4r^2\rr=0.
	\eeq
}
\end{itemize}

\section{Numerical Solution}
\label{B}
\begin{figure}[!h]
	\centering
	\includegraphics[width=16.3cm]{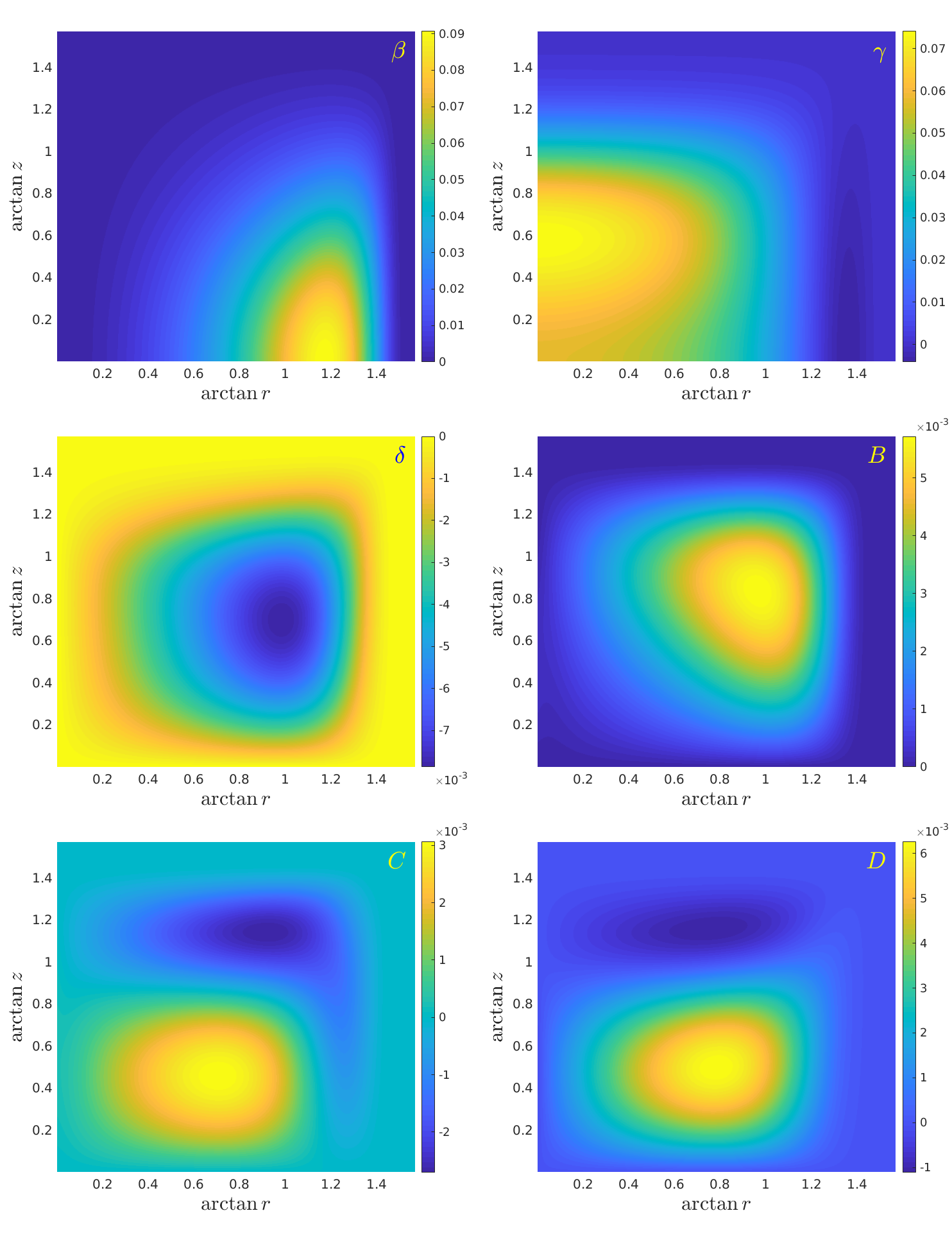}
	\caption{The numerical solution to the equations given in Appendix
		\ref{A}, part one. }
	\label{fig:sol_part1}
\end{figure}
\begin{figure}[!h]
	\centering
	\includegraphics[width=16.3cm]{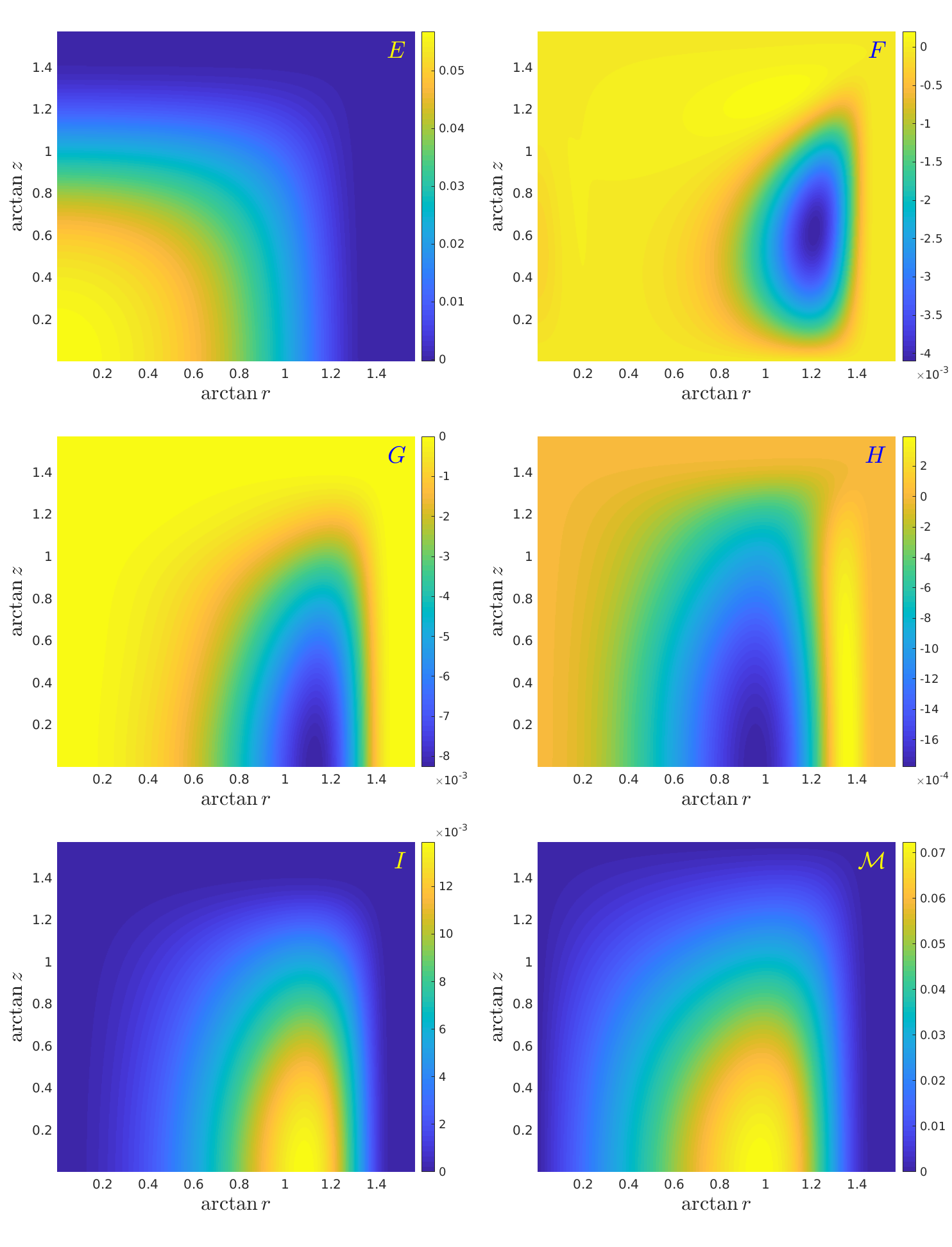}
	\caption{The numerical solution to the equations given in Appendix
		\ref{A}, part two. }
	\label{fig:sol_part2}
\end{figure}

We perform a change of coordinates
\beq
\mathsf{x}=\arctan r, \qquad
\mathsf{y}=\arctan z,
\eeq
and discretize the latter variables on an equidistant lattice of
$512^2$ points and use a fourth-order 5-stencil finite difference
scheme to calculate the derivatives.
We solve the 11 coupled PDEs using a custom built CUDA C code using
the relaxation method.
To this end, we calculate the solutions for each source term (the
latter terms in each of the equations in Appendix \ref{B})
separately and add the resulting solutions to get a final solution for
the fields $\beta$, $\gamma$, $\delta$, $B$, $C$, $D$, $E$, $F$, $G$,
$H$, and $I$.
Using this solution, we check that the total solution is still
satisfying the full system of equation and then we use Eq.~\eqref{eomM} to
calculate $\mathcal{M}$ from which the EDM can be computed using
Eq.~\eqref{eq:DeltaDi}.
The solution is shown in Figs.~\ref{fig:sol_part1}, \ref{fig:sol_part2}.

\end{document}